\documentclass[11pt,graphicx]{article}
\usepackage{amssymb,amsmath,amsfonts}
\usepackage{graphicx}
\usepackage{graphics}
\usepackage{eepic,epsfig}
\begin{document}
\title{Induced self-energy on a static scalar charged particle in the spacetime of a global monopole with finite core}
\author{D. Barbosa{\thanks{E-mail: denis.barros@ifpb.edu.br}}, U. de Freitas\thanks{E-mail: umbelino@fisica.ufpb.br}\\ and 
E. R. Bezerra de Mello\thanks{E-mail: emello@fisica.ufpb.br}\\\\
Instituto Federal de Educa\c c\~{a}o, Ci\^{e}ncia e Tecnologia da Para\'{\i}ba\\
58.800-970, Sousa, PB, Brazil\\
Departamento de F\'{\i}sica-CCEN, Universidade Federal da Para\'{\i}ba\\
58.059-970, C. Postal 5.008, J. Pessoa, PB,  Brazil}
\maketitle

\begin{abstract}
We analyze the induced self-energy and self-force on a scalar point-like charged test particle placed at rest in the spacetime of a global monopole admitting a general spherically symmetric inner structure to it. In order to develop this analysis we calculate the three-dimensional Green function associated with this physical system. We explicitly show that for points outside the monopole's core the scalar self-energy presents two distinct contributions. The first one is induced by the non-trivial topology of the global monopole considered as a point-like defect and the second is a correction induced by the non-vanishing inner structure attributed to it. For points inside the monopole, the self-energy also present a similar structure, where now the first contribution depends on the geometry of the spacetime inside. As illustrations of the general procedure adopted, two specific models, namely flower-pot and the ballpoint-pen, are considered for the region inside. For these two different situations, we were able to obtain exact expressions for the self-energies and self-forces in the regions outside and inside the global monopole.
\end{abstract}
\section{Introduction}
\label{Int}
Global monopoles are heavy spherically symmetric topological objects which may have been formed by the vacuum phase transition in the early Universe after Planck time \cite{Kibble,V-S}. Although the global monopole was first introduced by Sokolov and Starobinsky in \cite{Soko}, its gravitational effects has been analyzed by Barriola and Vilenkin \cite{BV}. In the latter it is shown that for points far away from the monopole's center, the geometry of the spacetime can be given by the line element below:
\begin{equation}
ds^2=-dt^2+dr^2+\alpha^2r^2(d\theta^2+\sin^2\theta d\varphi^2)\ ,  
\label{gm}
\end{equation}
where the parameter $\alpha^2$, smaller than unity, depends on the energy scale $\eta$ where the phase transition spontaneously occur. The spacetime described by (\ref{gm}) has a non-vanishing scalar curvature, $R=\frac{2(1-\alpha^2)}{\alpha^2r^2}$, and presents a solid angle deficit $\delta\Omega= 4\pi^2(1-\alpha^2)$. 

Although the geometric properties of the spacetime outside the monopole are very well understood, there are no explicit expressions for the components of the metric tensor in the region inside. \footnote{Considering the model described in \cite{BV}, the expressions for the metric tensor in the region inside the monopole, are expressed in terms of nonsolvable integral equation involving the energy-momentum tensor associated with the iso-scalar field, $\varphi^a$, which on the other hand depends on the components of the metric itself \cite{Mello1}.} As a consequence of this fact, many interesting investigations of physical effects associated with global monopole consider this object as a point-like defect. Adopting this simple model, calculations of vacuum polarization effects associated with bosonic \cite{Lousto} and fermionic quantum fields \cite{Mello2}, in four-dimensional global monopole spacetime, present divergence on the monopole's core. Moreover, considering higher-dimensional spacetime, vacuum polarization effects associated with bosonic \cite{Mello2a} and fermionic \cite{Mello2b} quantum fields, also present divergences on the monopole's core.

A very well known phenomenon that occur with an electric charged test particle placed at rest in a curved spacetime, is that it may become subjected to an electrostatic self-interactions. The origin of this induced self-interaction resides on the non-local structure of the field caused by the spacetime curvature and/or non-trivial topology. This phenomenon has been analyzed in an idealized cosmic string spacetime by Linet \cite{Linet} and Smith \cite{Smith}, independently, and also in the spacetime of a global monopole considered as a point-like defect in \cite{Mello3}. In these analysis, the corresponding self-forces are repulsive and depend on the square of electric charge; moreover they present divergences on the respective defects' core. A possible way to circumvent the divergence problem is to consider these defects as having a non-vanishing radius, and attributing for the region inside a structure. For the cosmic string, two different models have been adopted to describe the geometry inside it: the ballpoint-pen model proposed independently by Gott and Hiscock \cite{Gott}, replaces the conical singularity at the string axis by a constant curvature spacetime in the interior region, and flower-pot model \cite{BA}, presents the curvature concentrated on a ring with the spacetime inside the string been flat.  Khusnutdinov and Bezerra in \cite{NV}, revisited the induced electrostatic self-energy problem considering the Hiscock and Gott model for the region inside the string. As to the global monopole the electrostatic self-energies problem have been analyzed considering for the region inside, the flower-pot model in \cite{Mello4} and ballpoint pen in \cite{Barbosa}. In both analysis it was observed that the corresponding self-forces are proportional to the square of electric charge and are finite at the monopole's core center.\footnote{The analysis of vacuum polarization effect associated with scalar and fermionic fields, considering the flower-pot model for the region inside the monopole, have been developed in \cite{Mello5} and \cite{Mello6}, respectively.} Differently from the electrostatic analysis, the induced self-energy on scalar charged point-like particles on a curved spacetime reveals peculiarities \cite{Burko,Wiseman} due to the nonminimal curvature coupling with the geometry. In the case of of Schwarzschild spacetime, the scalar self-force on a scalar charged particle at rest vanishes for minimal coupling \cite{Zelnikov}. The self-energy on scalar particle on the wormhole spacetime has been developed recently in \cite{NV1}. In the latter, the authors have observed that the self-force is zero for conformal curvature coupling, $\xi=1/8$, for massless field.

In this paper we decide to the analysis the induced self-energy and self-force on a scalar point-like charged test particle placed at rest in the global monopole spacetime considering a inner structure to it. In fact, as first part of this investigation, we shall develop a general analysis admitting that the monopole has a non-vanishing radius, $a$, and a spherically symmetric inner structure. As two different applications of this formalism we shall adopt the flower-pot and ballpoint pen models for the region inside. The organization of this paper is the following: In section \ref{Self} we present our approach to consider the geometry of the spacetime under consideration and present the relevant field equations associated with the scalar charged particle. The explicit formula for the self-energy will be also presented. As illustrations of the general analysis, in section \ref{Applic} the flower-pot and ballpoint pen models for the inner region will be considered separately. For both cases we analyze the behavior of the self-energy in various asymptotic regions of the parameter. In section \ref{Conc}, we present our conclusions and more relevant remarks. We leave for the appendices \ref{appA} and \ref{appB}, specific details related with the renormalizations Green functions in point-like global monopole spacetime and in the ballpoint pen model. 

\section{Self-energy}
\label{Self} 
In this section we shall develop our approach to investigate the scalar self-energy in the global monopole spacetime, admitting a general spherically symmetric spacetime for its inner region.

\subsection{The model}
Many investigations concerning physical effects around a global monopole are developed considering it as idealized point-like defect. In this case the geometry of the spacetime is described by line element (\ref{gm}) for all values of the radial coordinate. However, a realistic global monopole has a characteristic core radius; for example, considering the model proposed by Barriola and Vilenkin \cite{BV}, the line element given by (\ref{gm}) is attained for the radial coordinate much lager than its characteristic core radius, which depends on the inverse of the energy scale where the global $O(3)$ symmetry is spontaneously broken to $U(1)$. Explicit expressions for the components of the metric tensor in whole space have not yet been found. Here, in this paper we shall not go into the details about this calculation. Instead, we shall consider a simplified model described by two sets of the metric tensor for two different regions, continuous at a spherical shell of radius $a$. In the exterior region corresponding to $r>a$, the line element is given by (\ref{gm}), while in the interior region, $r<a$, the geometry is described by the static spherically symmetric line element
\begin{equation}
ds^{2}=-dt^{2}+v^{2}(r)dr^{2}+w^{2}(r)(d\theta ^{2}+\sin ^{2}\theta d\varphi ^{2}) \ .  
\label{gm1}
\end{equation}
Because the metric tensor must be continuous at the boundary of the core, the functions $v(r)$ and $w(r)$ must satisfy the conditions
\begin{equation}
v(a)=1 \ {\rm and} \ w(a)=\alpha a \ .  \label{bound}
\end{equation}
The flower-pot and ballpoint pen models for a global monopole, analyzed in \cite{Mello4} and \cite{Barbosa}, respectively, can be obtained from the general expression given in (\ref{gm1}) by taking 
\begin{eqnarray}
v(r)=1 \ , \ w(r)=r+(\alpha-1)a 
\label{FP}
\end{eqnarray}
for the flower-pot model and
\begin{eqnarray}
v(r)=\alpha\left[1-r^2/a^2(1-\alpha^2)\right]^{-1/2} \ , \ w(r)=\alpha r \ ,
\label{BP}
\end{eqnarray}
for the ballpoint pen one.

\subsection{Scalar self-energy}
The action associated with a massive scalar field, $\phi$, coupled with a scalar charge density, $\rho$, in a curved background spacetime can be given by 
\begin{eqnarray}
\label{Action}
S=-\frac12\int \ d^4x \ \sqrt{-g} \ \left(g^{\mu\nu}\nabla_\mu\phi\nabla_\nu\phi+\xi R\phi^2+m^2\phi^2\right) +\int \ d^4x \ \sqrt{-g} \ \rho \ \phi \ ,
\end{eqnarray}
where the first part corresponds to the Klein-Gordon action with an arbitrary curvature coupling, $\xi$, and the second part contains the interaction term. In the above equation $R$ represents the scalar curvature and $g$ the determinant of $g_{\mu\nu}$. The field equation can be obtained by varying the action with respect to the field. This provides
\begin{eqnarray}
\label{EM}
\left(\Box-\xi R-m^2\right)\phi=-\rho \ .
\end{eqnarray}

The physical system which we shall analyze in this paper corresponds to a particle at rest, consequently there is no time dependence on the field. Moreover, for the metric spacetime under consideration the equation of motion above reduces to:
\begin{eqnarray}
\label{EM1}
\left(\nabla^2-\xi R-m^2\right)\phi=-\rho \ .
\end{eqnarray}

Taking the variation of (\ref{Action}) with respect to the metric tensor we obtain the energy-momentum tensor of the system:
\begin{eqnarray}
T_{\mu\nu}&=&\rho \ \phi \ g_{\mu\nu}+\nabla_\mu\phi\nabla_\nu\phi-\frac12g_{\mu\nu}\left(g^{\lambda\chi}\nabla_\lambda\phi\nabla_\chi\phi+m^2\phi^2 \right)\nonumber\\
&+&\xi\left(G_{\mu\nu}\phi^2+g_{\mu\nu}\Box\phi^2-\nabla_\mu\nabla_\mu\phi^2\right) \ ,
\end{eqnarray}
$G_{\mu\nu}$ being the Einstein tensor.

The energy associated with the scalar particle reads
\begin{eqnarray}
E=\int \ d^3x \ \sqrt{-g} \ T^0_0 \ .
\end{eqnarray}

For static fields configurations and by using the motion equation (\ref{EM1}), this energy reads:
\begin{eqnarray}
\label{SE}
E=\frac12\int \ d^3x \ \sqrt{-g} \  \rho \ \phi \ .
\end{eqnarray}
Moreover, (\ref{SE}) can be expressed in terms of a bilinear function of charge density by using the three-dimensional Green function defined by the  differential operator in (\ref{EM1}):
\begin{eqnarray}
\label{Green}
\left(\nabla^2-\xi R-m^2\right)G({\vec{x}},{\vec{x}}')=-\frac{\delta^3({\vec{x}}-{\vec{x}}')}{\sqrt{-g}} \ .
\end{eqnarray}
This equation becomes:
\begin{eqnarray}
\label{E1}
E=\frac12\int\int \ d^3x \sqrt{-g(x)} \ d^3x' \sqrt{-g(x')} \  \rho({\vec{x}}) G({\vec{x}},{\vec{x}}') \ \rho({\vec{x}'}) \ .
\end{eqnarray}

Considering now a point-like scalar charge at rest at the point $x_p$, the charge density takes the form,
\begin{eqnarray}
\label{Charge}
\rho(x)&=&q\int_{-\infty}^\infty\frac{d\tau}{\sqrt{-g}}\delta^4(x-x_p(\tau)) \nonumber\\
\rho(x)&=&\frac{q}{\sqrt{-g}}\delta^3({\vec{r}}-{\vec{r}}_p) \ .
\end{eqnarray}

Finally substituting (\ref{Charge}) into (\ref{E1}), we obtain for the energy the following expression:
\begin{eqnarray}
\label{E2}
E=\frac{q^2}2G({\vec{r}_p},{\vec{r}_p})  \ .
\end{eqnarray}

The evaluation of the Green function that we need for the calculation of the energy is divergent at the coincidence limit. There are several methods to obtain a finite result. By the approach suggested by Quinn in \cite{Quinn}, one obtains a finite expression by using the {\it Comparison Axiom}.\footnote{The comparison approach has also be used by Quinn and Wald \cite{Q-W} in the calculation of electromagnetic and gravitational self-energy.}  Here we shall use a general renormalization approach in curved spacetime \cite{B-D}, which consists to subtract from the Green function its DeWitt-Schwinger asymptotic expansion. Although we have started with a system in a four-dimensional spacetime, it becomes effectively defined in the three-dimensional section of the global monopole spacetime. In general there are two types of divergences in the expansion of the Green function, namely, pole and logarithmic ones. In three-dimensional case which we are interested in, there is only pole divergence. Following the general procedure given in \cite{Chris}, the singular behavior of the three-dimensional Green function associated with a massive scalar field reads:
\begin{equation}
G_{Sing}(x,x')=\frac1{4\pi}\left[\frac1{\sqrt{2\sigma}}-m\right]+O(\sigma) \ . \label{Had3D}
\end{equation}
Adopting the renormalizion approach for curved space, the scalar self-energy is given by
\begin{eqnarray}
\label{ERen}
E_{Ren}&=&\frac{q^2}2G_{Ren}({\vec{r}_p},{\vec{r}_p})  \  ,\nonumber\\
&=&\frac{q^2}2\lim_{{\vec{r}}\to{\vec{r}}_p}[G({\vec{r}_p},{\vec{r}})-G_{Sing}({\vec{r}}_p,{\vec{r}})]
\end{eqnarray}

\subsection{Green function}
Taking into account the spherical symmetry of the problem, the scalar Green function can be expressed by the following ansatz:
\begin{equation}
G(\vec{r},\vec{r}')=\sum_{l=0}^\infty\sum_{m=-l}^lg_l(r,r')Y_l^m(\theta ,\varphi )Y_l^{m\ast}(\theta',\varphi')\ ,  \label{Green-a}
\end{equation}
with $Y_l^m(\theta ,\varphi )$ being the ordinary spherical harmonics. Substituting (\ref{Green-a}) into (\ref{Green}), considering the general expression for the metric tensor (\ref{gm1}) and using the well known closure relation for the spherical harmonics, we arrive at the following differential equation for the unknown radial function $g_l(r,r')$:
\begin{eqnarray}
\label{gr}
\left[\frac d{dr}\left(\frac{w^2}{v}\frac{d}{dr}\right)-l(l+1)v-\xi Rvw^2-m^2vw^2\right]g_l=-\delta(r-r') \ ,
\end{eqnarray}
with the Ricci scalar being given by
\begin{equation}
R=\frac2{w^2}+\frac{4v'w'}{wv^3}-\frac{4w''}{wv^2}-\frac{2(w')^2}{w^2v^2} \ . 
\end{equation}
For $\xi=0$ and $m=0$ the differential equation above is similar to that one obtained for the analysis of electrostatic self-energy in \cite{Mello4}. Moreover; as to the solution of (\ref{gr}) some comments deserves to be mentioned: 
\begin{itemize}
\item Because the function $v(r)$ and $w(r)$ are continuous at $r=a$, it follows that $g_l(r,r')$ should be continuous at this point; however, due to the second radial derivative of the function $w(r)$ in the Ricci scalar, a Dirac-delta function contribution on the Ricci scalar takes place if the first derivative of this function is not continuous at the boundary.\footnote{The presence of a Dirac-delta contribution on the Ricci scalar will imply a non-vanishing surface energy-momentum tensor located at the boundary.} (For the flower-pot model this fact occur and has been considered in \cite{Mello4}.) Naming by $\check{R}=\bar{R}\delta(r-a)$ the Dirac-delta contribution of the Ricci scalar, the junction condition on the boundary is:
\begin{equation}
\frac{dg_l(r)}{dr}|_{r=a^+}-\frac{dg_l(r)}{dr}|_{r=a^-}=\xi\bar{R} \ g_l(a) \ . \label{Cond1}
\end{equation}
\item The function $g_l(r)$ is continuous at $r=r'$, however by integrating (\ref{gr}) about this point, the first radial derivative of this function obeys the junction condition below: 
\begin{equation}
\frac{dg_l(r)}{dr}|_{r=r'^+}-\frac{dg_l(r)}{dr}|_{r=r'^-}=-\frac{v(r')}{w^2(r')} \ . \label{Cond2}
\end{equation}
\end{itemize}

Now after this general discussion, let us analyze the solutions of the homogeneous differential equation associated with (\ref{gr}) for regions inside and outside the monopole's core. Let us denote by $R_{1l}(r)$ and $R_{2l}(r)$ the two linearly independent solutions of the equation in the region inside; moreover, we shall assume that the function $R_{1l}(r)$ is regular at the core center $r=r_{c}$ and that the solutions are normalized by the Wronskian relation
\begin{equation}
R_{1l}(r)R_{2l}^{\prime }(r)-R_{1l}^{\prime }(r)R_{2l}(r)=-\frac{v(r)}{w^{2}(r)}.  \label{Wronin}
\end{equation}
In the region outside, the two linearly independent solution are:
\begin{equation}
\frac1{\sqrt{r}}I_{\nu_l}(mr) \ \ {\rm and} \ \frac1{\sqrt{r}}K_{\nu_l}(mr) \ ,
\end{equation}
where $I_\nu$ and $K_\nu$ are the modified Bessel functions of order
\begin{equation}
\nu_l=\frac1{2\alpha}\sqrt{(2l+1)^2+(1-\alpha^2)(8\xi-1)} \ .
\end{equation} 

Now we can write the function $g_l(r)$ as a linear combination of the above solutions with arbitrary coefficients for the regions $(r_{c}, \min (r',a))$, $(\min (r',a), \ \max (r',a))$, and $(\max (r',a), \infty)$. The requirement of the regularity at the core center and at the infinity reduces the number of these coefficients to four. These constants are determined by the continuity condition at the monopole's core boundary and at the point $r=r'$ by the junctions conditions given in (\ref{Cond1}) and (\ref{Cond2}), respectively. In this way we find the following expressions:
\begin{eqnarray}
g_l(r,r')&=&\frac{K_{\nu_l}(mr')R_{1l}(r)}{\alpha^2\sqrt{ar'}}\frac1{a{\cal{R}}_l^{(1)}(a)K_{\nu_l}(ma)-R_{1l}(a){\tilde{K}}_{\nu_l}(ma)}  \ \ {\rm for} \ r<a \ , {\label{gout-}}\\ 
g_l(r,r')&=&\frac{I_{\nu_l}(mr_<)K_{\nu_l}(mr_>)}{\alpha^2\sqrt{rr'}}-D_l^{(+)}(a)\frac{K_{\nu_l}(mr)K_{\nu_l}(mr')}{\alpha^2\sqrt{rr'}}\ \ {\rm for} \ r>a \ , \label{gout+}
\end{eqnarray}
in the case of the charged particle is outside the monopole, $r'>a$ , and 
\begin{eqnarray}
g_l(r,r')&=&R_{1l}(r_<)[R_{2l}(r_>)-D_l^{(-)}(a)R_{1l}(r_>)] \ \ {\rm for} \ r<a \ , \label{gin-} \\ 
g_l(r,r')&=&\frac{R_{1l}(r')K_{\nu_l}(mr)}{\alpha^2\sqrt{ar}}\frac1{a{\cal{R}}_l^{(1)}(a)K_{\nu_l}(ma)-R_{1l}(a){\tilde{K}}_{\nu_l}(ma)} \ \ {\rm for} \ \ r>a \ , \label{gin+}
\end{eqnarray}
in the case of the charged particle is inside the monopole, $r'<a$. In these formulas, $r_{<}=\min (r,r')$ and $r_{>}=\max (r,r')$, and we have used the notations:
\begin{eqnarray}
D_{l}^{(+)}(a)&=&\frac{a{\cal{R}}_{l}^{(1)}(a)I_{\nu_l}(ma)-R_{1l}(a){\tilde{I}}_{\nu_l}(ma)}{a{\cal{R}}_{l}^{(1)}(a)K_{\nu_l}(ma)-R_{1l}(a){\tilde{K}}_{\nu_l}(ma)}  \ , \label{D+} \\
D_l^{(-)}(a)&=&\frac{a{\cal{R}}_{l}^{(2)}(a)K_{\nu_l}(ma)-R_{2l}(a){\tilde{K}}_{\nu_l}(ma)}{a{\cal{R}}_{l}^{(1)}(a)K_{\nu_l}(ma)-R_{1l}(a){\tilde{K}}_{\nu_l}(ma)} \ . \label{D-}
\end{eqnarray}
For a given function  $F(z)$, we use the notation
\begin{eqnarray}
{\tilde{F}}(z)=zF'(z)-\frac12F(z) 
\end{eqnarray}
and for a solution $R_{jl}(r)$, with $j= 1, \ 2$,
\begin{equation}
\label{Rsing}
{\cal{R}}_l^{(j)}(a)=R_{jl}'(a)+\xi{\bar{R}}R_{jl}(a) \ .
\end{equation}
In the above definition $R'_{jl}(a)=\frac{dR_{jl}(r)}{dr}|_{r=a}$.

Before to go for a specific model, let us still continue the investigation of the self-energy for this general spherically symmetric spacetime. First we shall consider the case where the charge is outside the monopole's core. Substituting (\ref{gout+}) into (\ref{Green-a}) we see that the Green function is expressed in terms of two contributions:
\begin{equation}
G(\vec{r},\vec{r}')=G_{gm}(\vec{r},\vec{r}')+G_c(\vec{r},\vec{r}') \ , 
\end{equation}
where
\begin{eqnarray}
G_{gm}(\vec{r},\vec{r}')=\frac1{4\pi\alpha^2{\sqrt{rr'}}}\sum_{l=0}^\infty (2l+1)I_{\nu_l}(mr_<)K_{\nu_l}(mr_>)P_l(\cos\gamma) \label{G-gm}
\end{eqnarray}
and 
\begin{eqnarray}
G_c(\vec{r},\vec{r}')=-\frac1{4\pi\alpha^2{\sqrt{rr'}}}\sum_{l=0}^\infty (2l+1)D_l^{(+)}(a)K_{\nu_l}(mr') K_{\nu_l}(mr)P_l(\cos\gamma) \ . \label{Gc}
\end{eqnarray}
The first part corresponds to the Green function for the geometry of a point-like global monopole and the second is induced by the non-trivial structure of its core. In the formulas above, $\gamma$ is the angle between both directions $(\theta, \ \varphi)$ and  $(\theta', \ \varphi')$ and $P_l(x)$ represents the Legendre polynomials of degree $l$.

The induced scalar self-energy is obtained by taking the coincidence limit in the renormalized Green function. We can observe that for points with $r>a$, the core-induced term (\ref{Gc}) is finite and the divergence appears in the point-like monopole part only. So, in order to provide a finite and well defined result for (\ref{ERen}), we have to renormalize the Green function $G_{gm}(\vec{r},\vec{r}')$ only. First of all we may take $\gamma=0$ in the above expressions. The renormalized Green function is expressed by:
\begin{equation}
G_{Ren}(r_p,r_p)=G_{gm,ren}(r_p,r_p)+G_c(r_p,r_p) \ ,  \label{Gren}
\end{equation}
where 
\begin{equation}
G_{gm,ren}(r_p,r_p)=\lim_{r\to r_p}[G_{gm}(r,r_p)-G_{Sing}(r,r_p)] \ .
\end{equation}
For points outside the core, the radial one-half of geodesic distance becomes $|r-r'|/2$, we have
\begin{equation}
G_{Sing}(r,r')=\frac1{4\pi}\left[\frac1{|r-r'|}-m\right] \ .
\end{equation}

Now, by using $G_m(r',r)$ given in (\ref{G-gm}), we have:
\begin{equation}
G_{gm,ren}(r,'r)=\frac1{4\pi r'}\lim_{r\to r'}\left[\frac1{\alpha^2}\sum_{l=0}^\infty{(2l+1)}I_{\nu_l}(mr_<)K_{\nu_l}(mr_>)-\frac{1}{1-t}\right]+\frac{m}{4\pi} ,  \label{Gmren1}
\end{equation}
where $t=r_{<}/r_{>}$. In order to evaluate the limit on the right hand side of the above equation, we note that
\begin{equation}
\lim_{t\to 1}\left( \frac{1}{\alpha }\sum_{l=0}^{\infty }t^{l/\alpha+1/2\alpha -1/2}-\frac{1}{1-t}\right) =0.  
\label{RelLim}
\end{equation}
So, as consequence of this relation and replacing in (\ref{Gmren1}) the expression $1/(1-t)$ by the first term in the brackets in (\ref{RelLim}), we find\footnote{In appendix \ref{appA} we derive a different proof of the renormalized Green function ({\ref{Gmren2}}).}
\begin{equation}
G_{{gm,ren}}(r',r')=\frac{S_{(\alpha )}(mr')}{4\pi\alpha r}+\frac{m}{4\pi} \ ,
\label{Gmren2}
\end{equation}
with
\begin{equation}
S_{(\alpha)}(mr')=\sum_{l=0}^\infty\left[\frac{(2l+1)}{\alpha}I_{\nu_l}(mr')K_{\nu_l}(mr')-1\right] \label{S} \ .
\end{equation}

Two special situations deserve to be analyzed:
\begin{itemize}
\item For the case where $\xi=0$, we have $\nu_l=\frac1{2\alpha}{\sqrt{4l(l+1)+\alpha^2}}$. Taking the limit $m\to 0$ in (\ref{S}), using \cite{Grad,Abra} we get a position independent expression named,
\begin{equation}
S_{\alpha}=\sum_{l=0}^\infty\left[\frac{2l+1}{\sqrt{4l(l+1)+\alpha^2}}-1\right] \ .
\end{equation}
This expression coincides with the similar one obtained in the calculation of the electrostatic self-energy for an electric charge at rest in the point-like global monopole spacetime analyzed in \cite{Mello3}. The corresponding self-energy becomes a positive quantity for $\alpha<1$ and negative quantity for $\alpha>1$.
\item For $\xi=1/8$ we have $\nu_l=\frac{1}{2\alpha}({2l+1})$. Taking the limit $m\to 0$ in (\ref{S}) we see that the term inside the bracket vanishes, consequently $G_{{gm,ren}}(r',r')=0$ and the only contribution to the scalar self-energy comes from the core-induced part (\ref{Gc}). In fact under these circumstance, the differential equation obeyed by the Green function in the region outside the monopole is conformally related with the corresponding one in a flat space due to the conformal flatness of the space section of this metric tensor:
\begin{equation}
d{\vec{l}}^2=dr^2+\alpha^2r^2d\Omega_{(2)}=\rho^\lambda(d\rho^2+\rho^2d\Omega_{(2)}) \ ,
\end{equation}
with $\rho=(\alpha r)^{1/\alpha}$ being $\lambda=2(\alpha-1)$. Moreover, by explicit calculation we can show that $G_{gm}(\vec{r},\vec{r}')=\rho^{-\lambda/4}G_M (\vec{\rho},\vec{\rho}')\rho'^{-\lambda/4}$.
\end{itemize}

Finally the complete expression for scalar self-energy reads
\begin{eqnarray}
E_{Ren}=\frac{q^2}{8\pi\alpha r_p}S_{(\alpha)}(mr_p)+\frac{q^2m}{8\pi}-\frac{q^2}{8\pi\alpha^2r_p} \sum_{l=0}^\infty(2l+1)D_l^{(+)}(a)\left(K_{\nu_l}(mr_p)\right)^2 \ . \label{TE}
\end{eqnarray}

Before to start the investigation about the convergence of the core-induced part on the scalar self-energy, we would like to understand better the contribution due to the point-like global monopole only, given by
\begin{eqnarray}
E_{Ren}=\frac{q^2}{8\pi\alpha r_p}S_{(\alpha)}(mr_p)+\frac{q^2m}{8\pi} \ . \label{TE0}
\end{eqnarray}
In figure $1$ we compare the behavior of (\ref{TE0}) as function of $mr$ with the massless limit case analyzed in \cite{Mello3}, for two different values of the curvature coupling constant. Self-energies for massive fields are represented by dash lines and for massless fields by solid lines. In the left panel we consider $\xi=0$ and in the right panel $\xi=1$. In both case we adopted $\alpha=0.9$. By these graphs we can see that for massive fields the self-energy goes to zero faster than the massless ones; moreover we also can observe that the sign of the self-energy for massive fields depends crucially with the curvature coupling constant.
\begin{figure}[tbph]
\begin{center}
\label{fig1}
\begin{tabular}{cc}
\epsfig{figure=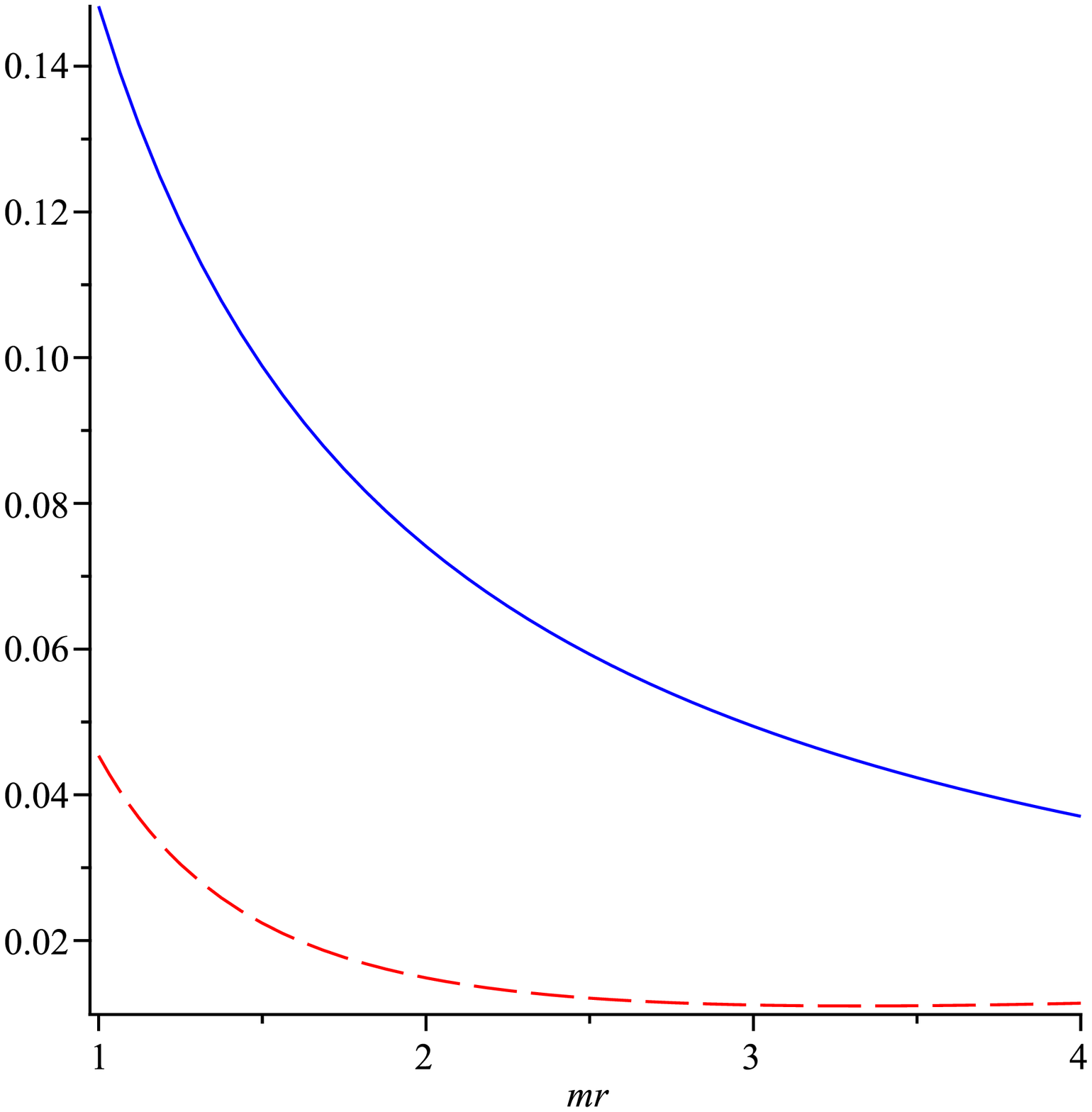, width=6.5cm, height=6.5cm,angle=0} & \quad 
\epsfig{figure=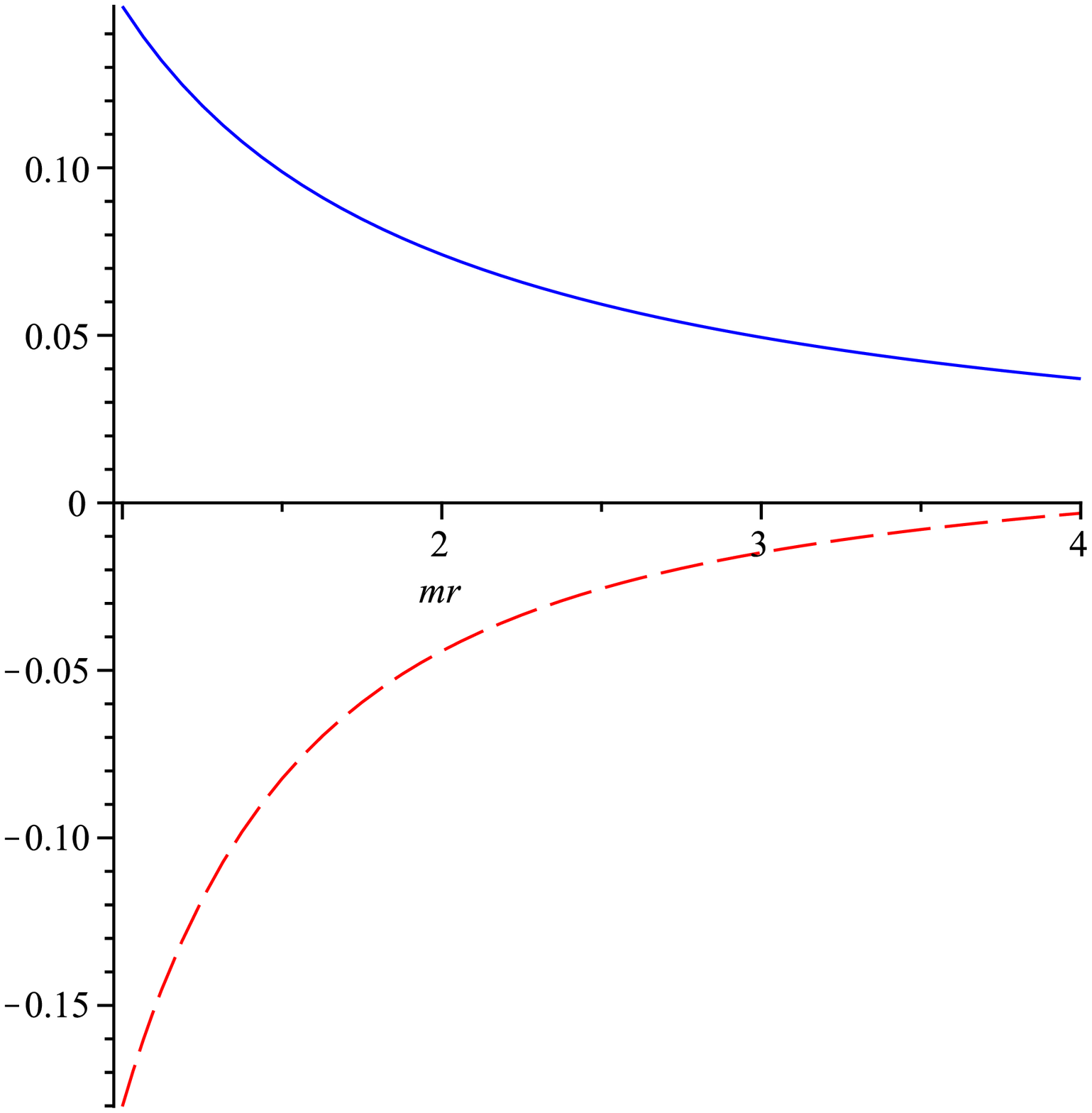, width=6.5cm, height=6.5cm,angle=0}
\end{tabular}
\end{center}
\caption{These graphs provide the behavior of $\frac{8\pi E_{Ren}}{q^2m}$ as function of $mr$ considering massive fields in dash line and massless fields in solid lines. In the left panel we took $\xi=0$, and in the right panel $\xi=1$. For both cases we have adopted $\alpha=0.9$.}
\label{fig1}
\end{figure}

In figure $2$ we plot in the left panel the behavior of (\ref{TE0}), considering two different values for the masses of the fields: $m$ and $2m$. We can see that for larger mass the self-energy goes to zero faster. Moreover, we exhibit in the right panel the behavior of the self energy at the point $mr=1$ as function of the curvature coupling. Explicitly is shown the changing in the sign of the self-energy. Here also we adopted $\alpha=0.9$.
\begin{figure}[tbph]
\begin{center}
\label{fig1}
\begin{tabular}{cc}
\epsfig{figure=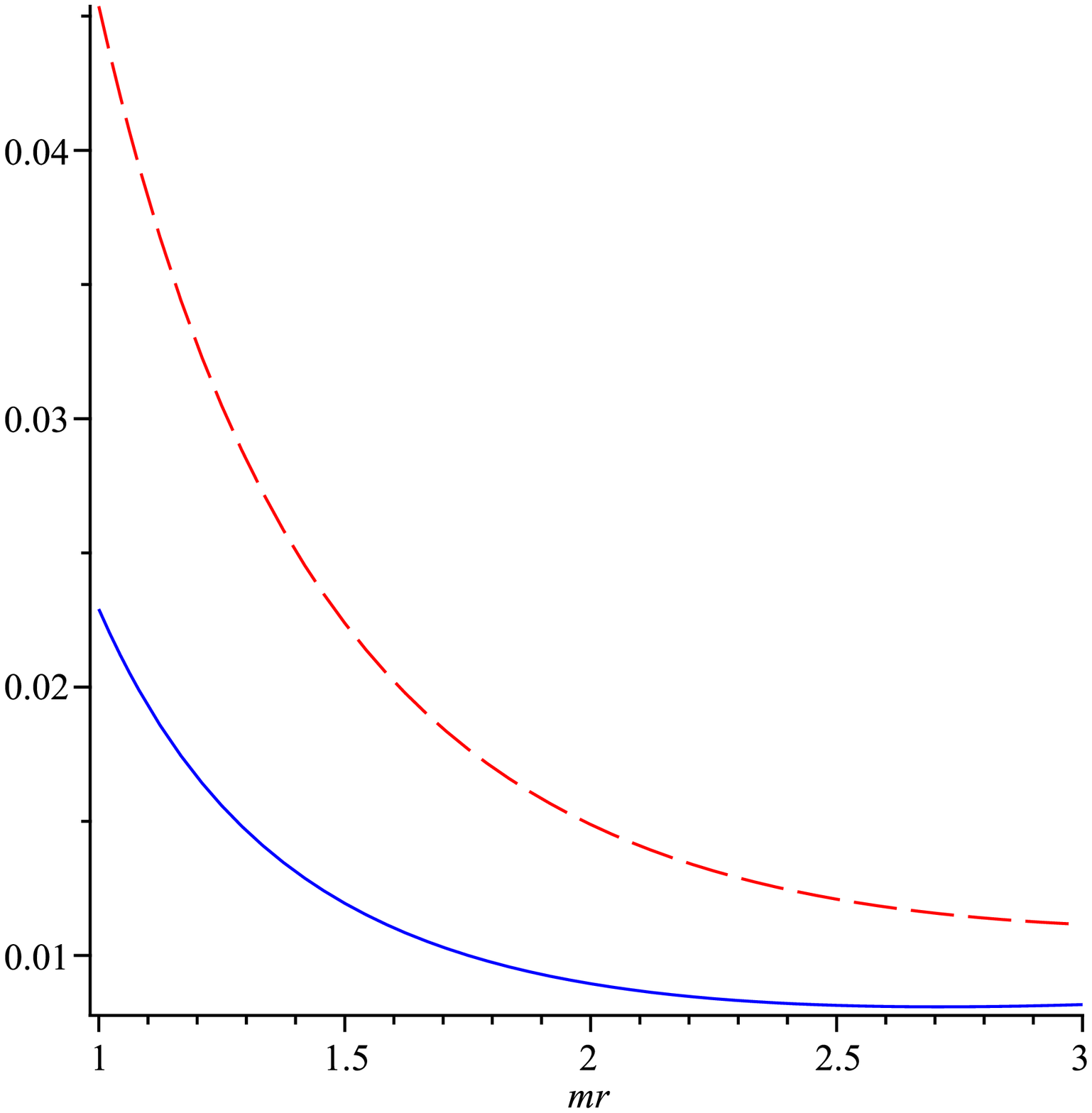, width=6.5cm, height=6.5cm,angle=0} & \quad 
\epsfig{figure=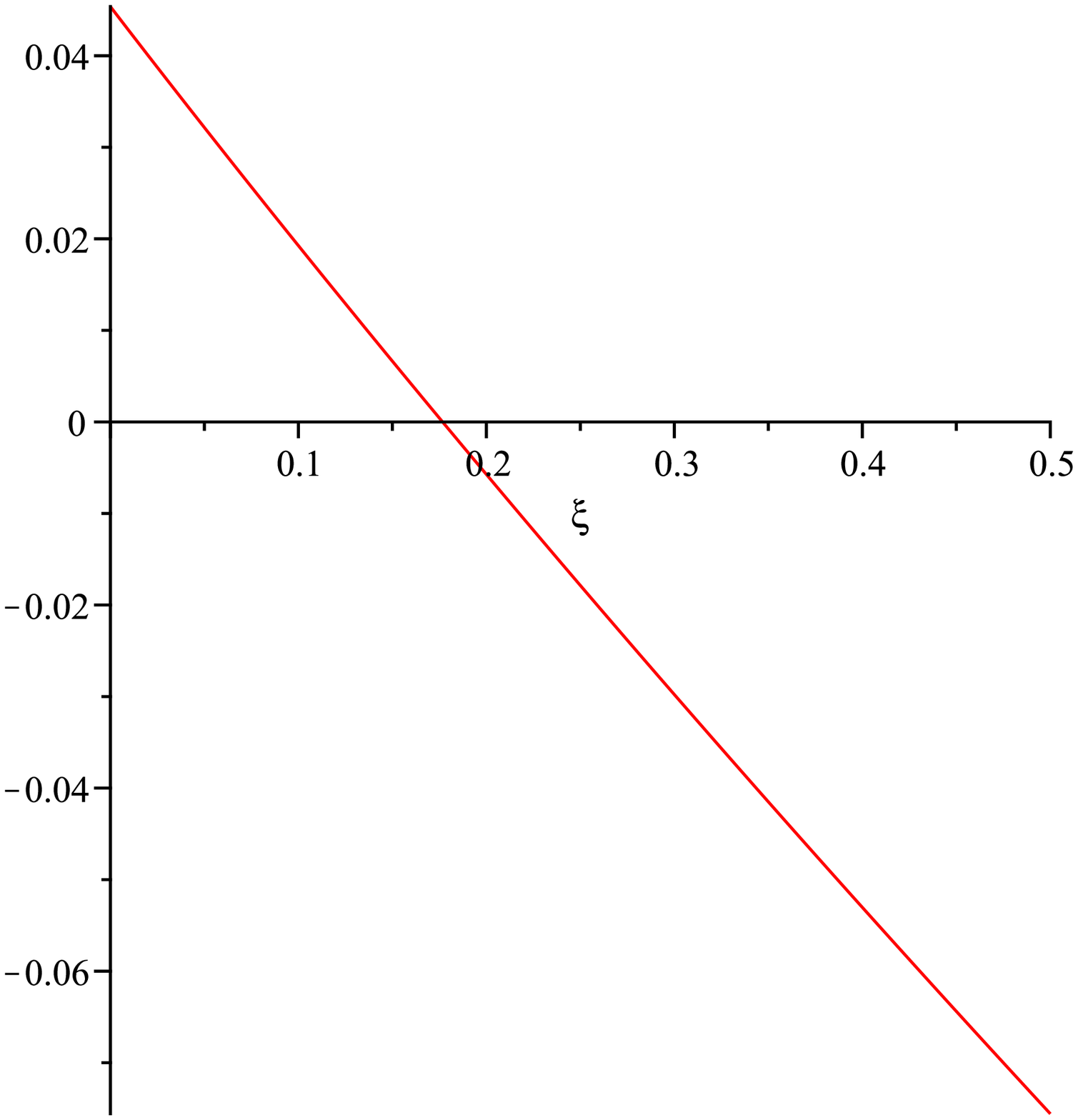, width=6.5cm, height=6.5cm,angle=0}
\end{tabular}
\end{center}
\caption{These graphs provide the behavior of $\frac{8\pi E_{Ren}}{q^2m}$. In the left plot it is exhibited this quantity as function of $mr$ for two different values of masses: the dash line is for mass $m$ and solid line for mass $2m$. In the right plot, we exhibit the behavior of the self-energy at the point $mr=1$ as function of $\xi$. In the latter we see explicitly that for specific value of $\xi$, the self-energy changes its sign. For both cases we adopted $\alpha=0.9$.}
\label{fig2}
\end{figure}

The convergence of the core-induced part on the scalar self-energy can be investigated by analyzing the general term inside the summation of the third term of (\ref{TE}) for large values of quantum number $l$, using the uniform asymptotic expansion for large orders of the modified Bessel functions $K_\nu$ and $I_\nu$ \cite{Abra}:
\begin{eqnarray}
\label{UAE}
K_\nu(\nu z)&\approx&\sqrt{\frac{\pi}{2\nu}}\frac{e^{-\nu\eta(z)}}{(1+z^2)^{1/4}}\sum_{k=0}^\infty(-1)^k\frac{u_k(w(z))}{\nu^k} \ , \nonumber\\
\ I_\nu(\nu z)&\approx&\frac1{\sqrt{2\pi\nu}}\frac{e^{\nu\eta(z)}}{(1+z^2)^{1/4}}\sum_{k=0}^\infty\frac{u_k(w(z))}{\nu^k} \ ,
\end{eqnarray}
where $w(z)=\frac1{\sqrt{1+z^2}}$ and $\eta(z)=\sqrt{1+z^2}+\ln\left(\frac{z}{1+{\sqrt{1+z^2}}}\right)$. The relevant expressions for $u_k(z)$ are: $u_0(w)=1$ and $u_1(w)=\frac{3w-5w^3}{24}$. In our expression $z=\frac{mr}{\nu}$. Keeping only the leading term in $1/\nu_l$ for the above expansions, we find 
\begin{eqnarray}
\left(K_{\nu_l}(mr)\right)^2\approx \frac\pi{2\nu_l}e^{-2\nu_l}\left(\frac{2\nu_l}{mr}\right)^{2\nu_l} 
\end{eqnarray}
and
\begin{eqnarray}
D_l^{(+)}(a)\approx -\frac{[R_{1l}(a)(2\nu_l-1)-2a{\cal{R}}_l^{(1)}(a)]}{\pi[R_{1l}(a)(2\nu_l+1)+ 2a{\cal{R}}_l^{(1)}(a)]}e^{2\nu_l}\left(\frac{ma}{2\nu_l}\right)^{2\nu_l} \ .
\end{eqnarray}
Substituting both results, the term inside the summation becomes
\begin{eqnarray}
\label{Approx}
-\alpha\frac{[R_{1l}(a)(2\nu_l-1)-2a{\cal{R}}_l^{(1)}(a)]}{[R_{1l}(a)(2\nu_l+1)+ 2a{\cal{R}}_l^{(1)}(a)]}\left(\frac ar\right)^{2\nu_l} \ ,
\end{eqnarray}
which provides a convergent result for $r>a$.

The self-force on a static test particle can be calculated by taking the negative gradient of the corresponding self-energy \cite{Burko},
\begin{equation}
\vec{f}=-{\vec{\nabla}}E_{Ren} \ .
\end{equation} 
Considering $\vec{f}=f_r\hat{r}$, the radial component of this force reads:
\begin{eqnarray}
f_r&=&\frac{q^2}{8\pi\alpha r_p^2}S_{(\alpha)}(mr_p)-\frac{q^2m}{8\pi\alpha^2r_p}\sum_{l=0}^\infty(2l+1)\left[\left(I_{\nu_l+1}(mr_p)+ \frac{\nu_l I_{\nu_l+1}(mr_p)}{mr_p}\right) K_{\nu_l}(mr_p)\right.\nonumber\\
&-&\left.I_{\nu_l}(mr_p)\left(K_{\nu_l+1}(mr_p)-\frac{\nu_lK_{\nu_l}(mr_p)}{mr_p}\right)\right]-\frac{q^2}{8\pi\alpha^2r^2_p} \sum_{l=0}^\infty(2l+1)D_l^{(+)}(a)\left(K_{\nu_l}(mr_p)\right)^2\nonumber\\
&-&\frac{q^2m}{4\pi\alpha^2r_p}\sum_{l=0}^\infty(2l+1)D_l^{(+)}(a)K_{\nu_l}(mr_p)\left(K_{\nu_l+1}(mr_p)-\frac{\nu_lK_{\nu_l}(mr_p)}{mr_p}\right) \ .
\end{eqnarray}

The second analysis that can be formally developed here, is related with the case when the charge is inside the core. The corresponding Green function can be written in the form
\begin{equation}
G(\vec{r},\vec{r}')=G_{0}(\vec{r},\vec{r}')+G_\alpha(\vec{r},\vec{r}') \ , 
\end{equation}
where
\begin{equation}
G_{0}(\vec{r},\vec{r}')=\frac{1}{4\pi }\sum_{l=0}^{\infty}(2l+1)R_{1l}(r_<)R_{2l}(r_>)P_{l}(\cos \gamma ) \ ,  \label{G0in}
\end{equation}
is the Green function for the background geometry described by the line element (\ref{gm1}) and the term
\begin{equation}
G_{\alpha}(\vec{r},\vec{r}')=-\frac{1}{4\pi }\sum_{l=0}^{\infty}(2l+1)D_l^{(-)}(a)R_{1l}(r)R_{1l}(r')P_{l}(\cos \gamma ) \ ,  \label{Galfain}
\end{equation}
is due to the global monopole geometry in the region $r>a$. For the points away from the core boundary the latter is finite in the coincidence limit.
The renormalized scalar self-energy for the charge inside is written in the form
\begin{equation}
\label{ERin}
E_{Ren}=\frac{q^2}2G_{0,Ren}(\vec{r}_p,\vec{r}_p)-\frac{q^{2}}{8\pi}\sum_{l=0}^{\infty }(2l+1)D_l^{(-)}(a)(R_{1l}(r_p))^2,
\end{equation}
where the renormalized Green function is given by
\begin{equation}
G_{0,Ren}(\vec{r}_p,\vec{r}_p)=\lim_{\vec{r}\to\vec{r}_p}\left[G_{0}(\vec{r},\vec{r}_p)-G_{Sing}(\vec{r},\vec{r}_p)\right] \  .  \label{G0renin}
\end{equation}
Because the divergent part of the Green function should have the same structure as (\ref{Had3D}), the above expression provides a finite result; moreover, notice that near the center of the core one has $R_{1l}(r)\approx\frac{I_{l+1/2}(m(r-r_{c}))}{\sqrt{r-r_c}}$ and the main
contribution into the second term on the right of (\ref{ERin}) comes from the term with $l=0$. Finally we can say that the self-force is again obtained by taking the negative gradient of (\ref{ERin}).

After we have developed the analysis for the scalar self-energy in the global monopole spacetime considering a general spherically symmetric inner structure to it, in the next section we shall apply this formalism for two different models for the geometry of region inside.

\section{Applications}
\label{Applic}
As we have mentioned in the Introduction, there is no closed expression for the metric tensor in the region inside the global monopole. In \cite{HL}, Harari and Lousto have proposed a simplified model for the monopole where the region inside the core is described by de Sitter geometry.\footnote{The vacuum polarization effects associated with a massless scalar field in the region outside the core of this model have been investigated in \cite{Mello7}.} Also for the cosmic string there is no closed expression for the metric tensor in the region inside; however in the literature there are two different models proposed to describe the geometry inside the core: the ballpoint-pen model proposed independently by Gott and Hiscock \cite{Gott} and flower-pot model \cite{BA}. Adopting the latter model for a global monopole, the calculations of vacuum polarization effects associated with massive scalar and fermionic fields have been developed in \cite{Mello5,Mello6}, respectively. Moreover, the analysis of electrostatic self-energies associated with electric charged particles placed at the rest in the global monopole spacetime considering both different models, have been calculated in \cite{Mello4,Barbosa}.
 
So, motived by these results we decided, to consider also both models, the flower-pot and ballpoint pen ones, in the analysis of the induced self-interaction associated with a scalar charge placed at rest in the global monopole spacetime.

\subsection{Flower-pot model}
\label{Flower}
The first model to be considered is the flower-pot one. For this model the interior line element has the form \cite{Mello4}
\begin{equation}
ds^{2}=-dt^{2}+dr^{2}+\left[ r+(\alpha -1)a\right] ^{2}(d^{2}\theta +\sin^{2}\theta d^{2}\varphi ) \ .  \label{intflow}
\end{equation}
In terms of the radial coordinate $r$ the origin is located at $r=r_{c}=(1-\alpha )a$. Defining $\tilde{r}=r+(\alpha -1)a$, the line element takes the standard Minkowskian form. As we have mentioned before, from the Israel matching conditions for the metric tensors corresponding to (\ref{gm}) and (\ref{intflow}), we find the singular contribution for the scalar curvature located on the bounding surface $r=a$ \cite{Mello5}:
\begin{equation}
\bar{R}=4\frac{(1-\alpha)}{\alpha a} \ .
\label{Rbar}
\end{equation}

In the region inside the global monopole the two linearly independent solutions for the radial functions are:
\begin{eqnarray}
R_{1l}(r)=\frac{I_{l+1/2}(m\tilde{r})}{\sqrt{\tilde{r}}} \ {\rm and} \ R_{2l}(r)=\frac{K_{l+1/2}(m\tilde{r})}{\sqrt{\tilde{r}}}  \ . \label{Flower-Int}
\end{eqnarray}
Having obtained the above solutions, the expressions for the Green functions in both, inside and outside regions, can be explicitly constructed, consequently the corresponding self-energies. These expressions depend on the coefficients $D^{(+)}_l(a)$ and $D^{(-)}_l(a)$, which can be explicitly provided as shown below:
\begin{eqnarray}
\label{D+-}
D^{(+)}_l(a)=\frac{n^{(+)}_l(a)}{d_l(a)}  \ \ {\rm and} \ \ D^{(-)}_l(a)=\frac{n^{(-)}_l(a)}{d_l(a)} \ ,
\end{eqnarray}
with
\begin{eqnarray}
\label{n+}
n^{(+)}_l(a)&=&I_{\nu_l}(ma)I_{l+1/2}(m\alpha a)\left[\frac{(l+4\xi(1-\alpha))}{\alpha}-\nu_l+\frac12\right]\nonumber\\
&+&ma\left[I_{\nu_l}(ma)I_{l+3/2}(m\alpha a) -I_{l+1/2}(m\alpha a)I_{\nu_l+1}(ma)\right] \ ,
\end{eqnarray}
\begin{eqnarray}
\label{n-}
n^{(-)}_l(a)&=&K_{\nu_l}(ma)K_{l+1/2}(m\alpha a)\left[\frac{(l+4\xi(1-\alpha))}{\alpha}-\nu_l+\frac12\right]\nonumber\\
&-&ma\left[K_{\nu_l}(ma)K_{l+3/2}(m\alpha a) -K_{l+1/2}(m\alpha a)K_{\nu_l+1}(ma)\right] 
\end{eqnarray}
and
\begin{eqnarray}
\label{d}
d_l(a)&=&K_{\nu_l}(ma)I_{l+1/2}(m\alpha a)\left[\frac{(l+4\xi(1-\alpha))}{\alpha}-\nu_l+\frac12\right]\nonumber\\
&+&ma\left[K_{\nu_l}(ma)I_{l+3/2}(m\alpha a) +I_{l+1/2}(m\alpha a)K_{\nu_l+1}(ma)\right] \ .
\end{eqnarray}

As the first analyse, let us consider the core-induced part of the self-energy for the region outside, Eq. ({\ref{TE}}), adopting specific values for the parameter $\xi$ and mass of the particle. Taking $\xi=0$, and by using \cite{Abra} to obtain the behavior of all functions contained in the coefficient defined in (\ref{D+-}), (\ref{n+}) and (\ref{d}), in the limit $m\to 0$, we have:
\begin{eqnarray}
D^{(+)}_l(a)\approx\frac{2}{\nu_l}\left(\frac{ma}2\right)^{2\nu_l}\frac{(\alpha+2l-2\alpha\nu_l)}{(2\alpha\nu_l+\alpha+2l)(\Gamma(\nu_l))^2} \ .
\end{eqnarray}
So the general term inside the summation of the core-induced part reads,
\begin{eqnarray}
\frac{\alpha(2l+1)}{\sqrt{4l^2+4l+\alpha^2}}\frac{4l(\alpha-1)}{(\sqrt{4l^2+4l+\alpha^2}+2l+\alpha)^2}\left(\frac ar\right)^{\sqrt{1+4l(l+1)/\alpha^2}} \ .
\end{eqnarray}
Which coincides with previous corresponding result found in \cite{Mello4}.

We can see that the core-induced part in ({\ref{TE}}) is divergent in near the boundary $r=a$. In order to verify this fact, we should analyze the general term in the summation for large value of $l$. Taking again the uniform asymptotic expansion for large orders of the modified Bessel functions in $D^{(+)}(a)$ and the same for the Macdonald function, $K_{\nu_l}(mr)$, we get,
\begin{eqnarray}
\label{Approx1}
-\frac{[\alpha-2\alpha\nu_l+2l+8\xi(1-\alpha)]}{[\alpha+2\alpha\nu_l+2l+8\xi(1-\alpha)]}\left(\frac ar\right)^{2\nu_l} \ .
\end{eqnarray}
At this point we want to mention that the contribution proportional to the curvature coupling, $\xi$, in the above expression is consequence of the delta-Dirac contribution in the Ricci scalar, given by $\xi R$, present in (\ref{Rsing}). 

For large value of $l$ we have $2\alpha\nu_l\approx(2l+1)+\frac{(1-\alpha^2)(8\xi-1)}{2(2l+1)}+O\left(\frac1{(2l+1)^2}\right)$. Substituting this expansion into (\ref{Approx1}), for the leading term in $1/l$, we obtain
\begin{eqnarray}
-\alpha(1-\alpha)\frac{(1-8\xi)}{4l}\left(\frac ar\right)^{2l/\alpha} \ .
\end{eqnarray}
Finally, after some intermediate steps we find:
\begin{equation}
\label{FP+}
E_{Ren}\approx-q^2\frac{(1-\alpha)(1-8\xi)}{32\pi\alpha a}\ln\left(1-\left(a/r_p\right)^{1/\alpha}\right) \ .
\end{equation}
We can see that the above result does not depend on the mass of the particle. In fact this happens because the leading order term in the expansions of the coefficient $D^{(+)}(a)$ there appear a power factor $(ma)^{2\nu_l}$, as to the square of the Macdonald function $K_{\nu_l}(mr)$, the leading power factor in the mass is $(mr)^{-2\nu_l}$. Moreover we can see from the above result that for $\xi=1/8$ there is no divergent contribution in the core-induced part, and that for $\xi>1/8 $ this contribution becomes negative.

Now let us turn our investigation of the self-energy for the region inside  the monopole. Substituting the functions (\ref{Flower-Int}) into the formulas (\ref{G0in}) and (\ref{Galfain}) for the corresponding Green function in the interior region one finds \cite{Abra},
\begin{eqnarray}
\label{G00}
G_{0}(\vec{r},\vec{r}')=\frac1{4\pi}\frac{e^{-mR}}{R} \ ,
\end{eqnarray}
with $R=\sqrt{(\tilde{r}')^2+(\tilde{r})^2-2\tilde{r}\tilde{r}'\cos\gamma}$, being $\gamma$ the angle between the two directions defined by the unit vectors $\hat{\tilde{r}}'$ and $\hat{\tilde{r}}$. Taking $\gamma=0$ we get $R=|r-r'|$. Because in the flower-pot model the geometry in the region inside the monopole is a Minkowski one, we have $G_{0}(\vec{r},\vec{r}')=G_{Sing}(\vec{r},\vec{r}')$, consequently $G_{0,Ren}(\vec{r}_p,\vec{r}_p)=0$. The scalar self-energy in this region is given only by the core-induced part:
\begin{equation}
\label{E-ind}
E_{Ren}=-\frac{q^2}{8\pi{\tilde{r}}_p}\sum_{l=0}^\infty(2l+1)D^{(-)}_l(a)\left(I_{l+1/2}(m\tilde{r}_p)\right)^2 \ ,
\end{equation}
being $\tilde{r}_p=r_p+(\alpha-1)a$. Near the core's center, $\tilde{r}_p\approx 0$, 
\begin{eqnarray}
I_{l+1/2}(m\tilde{r}_p)\approx\left(\frac{m\tilde{r}_p}{2}\right)^{l+1/2}\frac1{\Gamma(l+3/2)} \ , 
\end{eqnarray}
so the main contribution into the self-energy comes from the lowest mode, $l=0$, resulting in
\begin{equation}
\label{EE}
E_{Ren}\approx-\frac{q^2mD_0^{(-)}(a)}{4\pi^2} \ .
\end{equation}
Taking the expression for the coefficient $D_l^{(-)}(a)$, for $l=0$, and considering $\xi=0$, after some steps we find
\begin{equation}
D_0^{(-)}(a)=\frac12\frac{\pi e^{-m\alpha a}(\alpha-1)}{\sinh(m\alpha a)(\alpha+\alpha am-1)+\alpha am\cosh(m\alpha a)} \ .
\end{equation}
Finally substituting the above expression into (\ref{EE}), and taking the limit $m\to 0$ we obtain
\begin{equation}
E_{Ren}(r)\approx-\frac{q^2(\alpha-1)}{8\pi\alpha^2a} \ .
\end{equation}

Also we can calculate the core-induced contribution on the scalar self-energy near the boundary. Again, adopting a similar procedure as in the previous corresponding analysis, we can verify after some intermediate steps that the leading term inside the summation in (\ref{E-ind}) behaves as,
\begin{equation}
\frac1{4l}\frac{(1-\alpha)(8\xi-1)}{\alpha a}\left(\frac {\tilde{r}}{\alpha a}\right)^{2l} \ .
\end{equation}
Finally, taking this expression back into (\ref{E-ind}) we obtain,
\begin{equation}
\label{FP-}
E_{Ren}\approx-q^2\frac{(1-\alpha)(1-8\xi)}{32\pi\alpha a}\ln\left(1-\frac{\tilde{r}_p}{\alpha a}\right) \ .
\end{equation} 
Once more we can see that this divergent contribution vanishes for $\xi=1/8$. 

Although the above result has been obtained  under the condition $\frac{{\tilde{r}}_p}{\alpha a}< 1$, we may wonder about its value for the case where $a\to 0$. In this limit the monopole has no core and consequently the core-induced self-energy does not exist; on the other hand a direct application of this limit in (\ref{FP-}) provides a divergent result. A possible way to circumvent this problem and preserving the condition $\frac{{\tilde{r}}_p}{\alpha a}< 1$ is to take first the limit ${\tilde{r}}_p\to 0$. Accepting this procedure a vanishing self-energy is promptly obtained as intuitively expected.

Before to finish this application we want to cal attention that (\ref{FP+}) and (\ref{FP-}), for $\xi=0$, coincide, up the numerical factor $4\pi$, with the corresponding core-induced electrostatic self-energies derived in \cite{Mello4}.

In figure $3$ we exhibit the behavior of the renormalized self-energy for a charged test particle placed inside (left plot) and outside (right plot) the monopole's core as function of $mr$. In these analysis we consider minimal coupling ($\xi=0$), $\alpha=0.9$ and the product $ma=1$. 
\begin{figure}[tbph]
\begin{center}
\label{fig1}
\begin{tabular}{cc}
\epsfig{figure=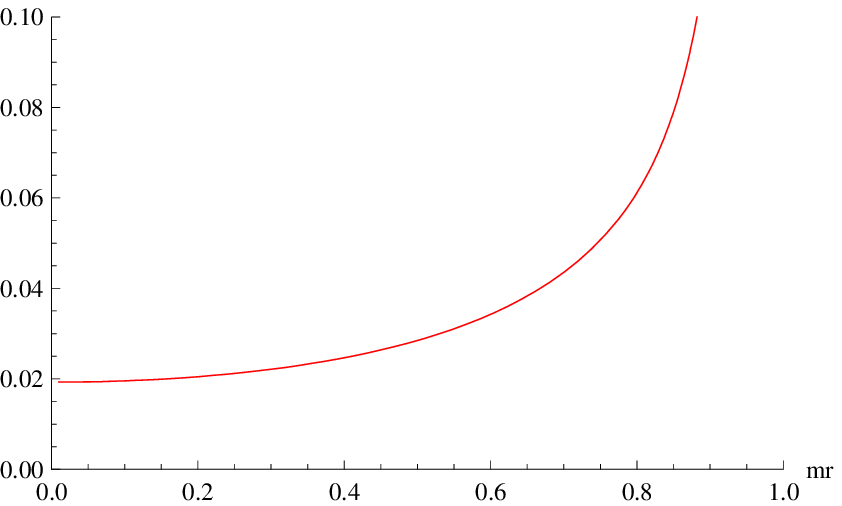, width=6.5cm, height=6.5cm,angle=0} & \quad 
\epsfig{figure=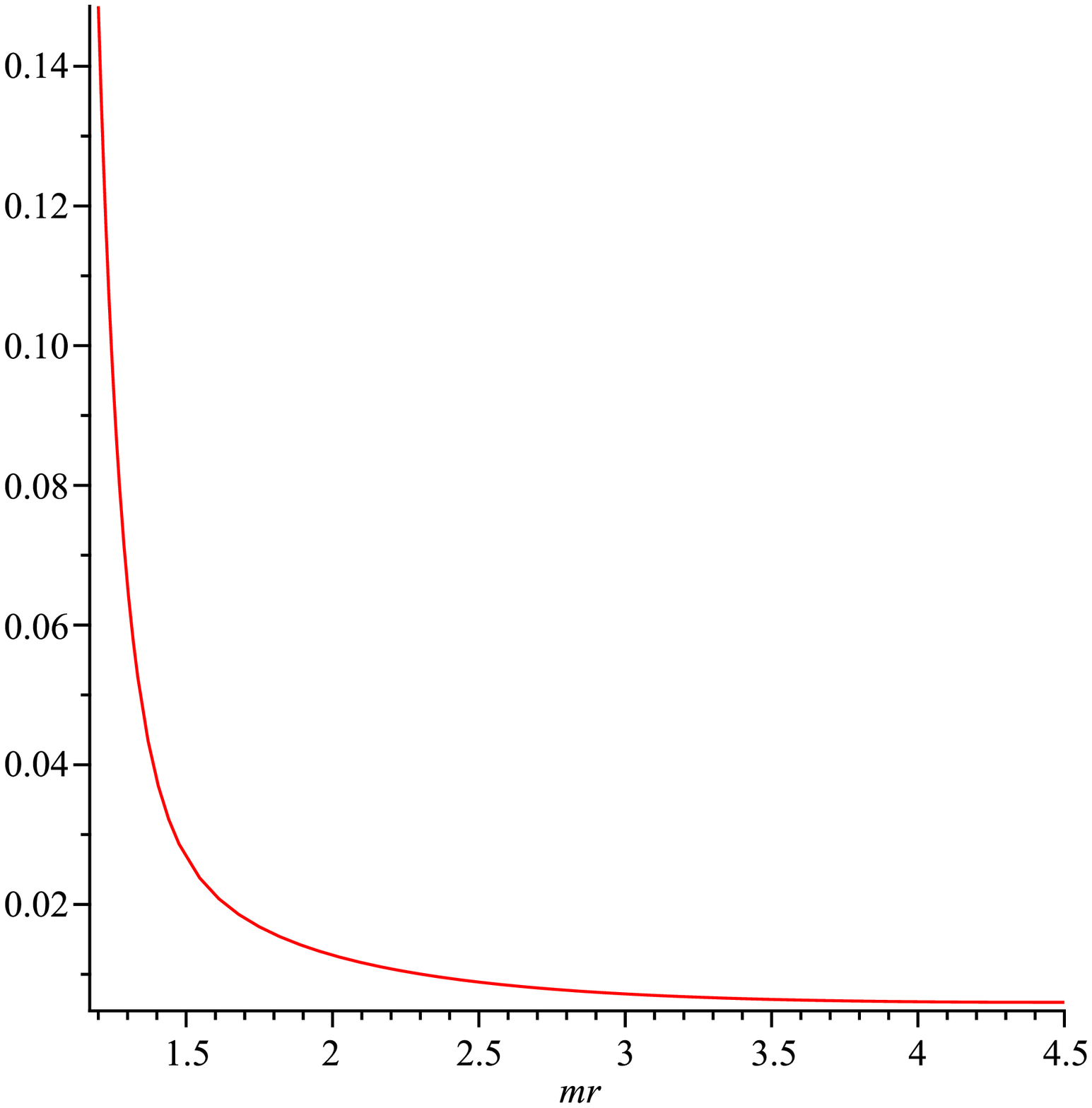, width=6.5cm, height=6.5cm,angle=0}
\end{tabular}
\end{center}
\caption{These graphs provide the behavior of $\frac{8\pi E_{Ren}}{q^2m}$ as function of $mr$ for the flower-pot model considering $\xi=0$ and $\alpha=0.9$. In the left plot we exhibit the behavior for the region inside the monopole, and in the right plot for the region outside the monopole.}
\label{fig2}
\end{figure}

\subsection{Ballpoint pen model}
\label{Ballpoint}
The ballpoint pen model has been used independently by Gott and Hiscock \cite{Gott} for describing the region inside the comic string's core. In this context the Dirac-delta singularity in the Ricci scalar is replaced by a constant curvature spacetime.

In previous publication \cite{Barbosa}, we have adapted this model for describing the region inside the global monopole. The inner region is covered by the coordinates $t\in(-\infty, \ \infty)$, $\rho\in[0, \ \rho_0)$, $\theta\in[0, \ \pi]$ and $\varphi\in[0, \ 2\pi]$. The metric tensor components are $C^1-$matched on the boundary, and there is no spherical surface stress energy. For the region inside, the metric has the following form:
\begin{eqnarray}
\label{BP1}
	ds^2=-dt^2+d\rho^2+\left(\frac{\rho_0}\epsilon\right)^2\sin^2\left(\frac{\epsilon\rho}{\rho_0}\right)(d\theta^2+\sin^2\theta d\varphi^2) \ .
\end{eqnarray}
The junction condition on the boundary with the external metric given by (\ref{gm}) provides,
\begin{eqnarray}
\alpha a=\frac{\rho_0}\epsilon\sin\epsilon \ , \ {\rm with} \ \ \alpha=\cos\epsilon \ .
\end{eqnarray}

In the inner region, the non zero components of the Riemann and Ricci tensors and scalar curvature are:
\begin{eqnarray}
	R^{\rho\theta}_{\rho\theta}=R^{\rho\varphi}_{\rho\varphi}=R^{\varphi\theta}_{\varphi\theta}=\frac{\epsilon^2}{\rho_0^2} \ , \ R^\rho_\rho=R_\theta^\theta =R^\varphi_\varphi=2\frac{\epsilon^2}{\rho_0^2} \ , \ {\rm and} \ R=6\frac{\epsilon^2}{\rho_0^2} \ .
\end{eqnarray}

In the cosmic string spacetime, Allen and Ottewill in \cite{BA}, have written the inner line element in terms of the continuation of the radial coordinate $r$. Here, this continuation is also adopted and the radial coordinate $r$ is related with $\rho$ coordinate by the relation,
\begin{eqnarray}
	\sin\left(\frac{\epsilon\rho}{\rho_0}\right)=\frac ra\sin\epsilon \ .
\label{rela}
\end{eqnarray}
In this case, the line element (\ref{BP1}) can be written as
\begin{eqnarray}
\label{gm2}
	ds^{2}=-dt^{2}+v^2(r)dr^{2}+\alpha^{2}r^{2}(d\theta^{2}+\sin^{2}\theta d\varphi^{2}) \ ,
\end{eqnarray}
where
\begin{equation}
v(r)=\frac\alpha{\sqrt{1-\frac{r^2}{a^2}\sin^2\epsilon}} \ . \label{v}
\end{equation}
In this coordinate the only non-vanishing Ricci tensor are:
\begin{eqnarray}
R^r_r=\frac{2v'(r)}{rv^3(r)} \ , \ R_\theta^\theta=R^\varphi_\varphi=\frac1{\alpha^2r^2}\frac{(\alpha^2rv'(r)+v^3(r)-\alpha^2v(r))}{v^3(r)} \ . 
\end{eqnarray}
As to the Ricci scalar, it reads
\begin{eqnarray}
R=\frac{4v'(r)}{rv^3(r)}+\frac2{\alpha^2r^2}-\frac{2}{r^2v^2(r)}=\frac{6(1-\alpha^2)}{\alpha^2a^2} \ .
\end{eqnarray}

By using this coordinate system, the two independent solutions for the homogeneous radial differential equation associated with (\ref{gr}), compatible with the Wronskian normalization (\ref{Wronin}), are
\begin{eqnarray}
\label{RBP}
R_{1l}(r)=\frac1{\sqrt{\alpha\kappa_l r}}P_\chi^{-l-1/2}(x_r) \ \ {\rm and} \ \ R_{2l}(r)=\frac1{\sqrt{\alpha\kappa_l r}}P_\chi^{l+1/2}(x_r) \ ,
\end{eqnarray}
being $P_\mu^\nu(x)$ the associated Legendre function and\footnote{By the property \cite{Abra} $P_\nu^\mu(x)=P_{-\nu-1}^\mu(x)$, the functions $R_{1l}$ and $R_{2l}$ are real ones even for complex value of $\chi$.}
\begin{eqnarray}
x_r=\sqrt{1-(r^2/a^2)\sin^2\epsilon} \ \ , \ \ \chi=-\frac12+\frac{\sqrt{1-[1+(m^2+\xi R)a^2]\alpha^2}}{\sin\epsilon} \ \ , \ \ \kappa_l=\frac2\pi(-1)^l \ .
\end{eqnarray}

Having obtained the above radial functions, the next step is the calculation of the Green functions for both regions of the space. The corresponding functions depend on the coefficients $D_l^{(+)}(a)$ and $D_l^{(-)}(a)$ given by
\begin{eqnarray}
\label{DBP+-}
D^{(+)}_l(a)=\frac{n^{(+)}_l(a)}{d_l(a)}  \ \ {\rm and} \ \ D^{(-)}_l(a)=\frac{n^{(-)}_l(a)}{d_l(a)} \ ,
\end{eqnarray}
where, in this case, 
\begin{eqnarray}
\label{nBP+}
n^{(+)}_l(a)&=&P_\chi^{-l-1/2}(\alpha)[(\chi-\nu_l)I_{\nu_l}(ma)-maI_{\nu_l+1}(ma)]\nonumber\\
&-&\frac{(\chi-l-1/2)}{\alpha}I_{\nu_l}(ma)P_{\chi-1}^{-l-1/2}(\alpha) \ , 
\end{eqnarray}
\begin{eqnarray}
\label{nBP-}
n^{(-)}_l(a)&=&K_{\nu_l}(ma)P_\chi^{l+1/2}(\alpha)(\chi-\nu_l)-K_{\nu_l}(ma)P_{\chi-1}^{l+1/2}(\alpha)\frac{(\chi+l+1/2)}\alpha\nonumber\\
&+&maP_\chi^{l+1/2}(\alpha)K_{\nu_l+1}(ma) \ , 
\end{eqnarray}
and
\begin{eqnarray}
\label{dBP}
d_l(a)&=&P_\chi^{-l-1/2}(\alpha)[(\chi-\nu_l)K_{\nu_l}(ma)+maK_{\nu_l+1}(ma)]\nonumber\\
&-&\frac{(\chi-l-1/2)}\alpha K_{\nu_l}(ma)P_{\chi-1}^{-l-1/2}(\alpha) \ .
\end{eqnarray}

As we have observed for the flower-pot model, the singular behavior of the core-induced part of the renormalized self-energy was given by analyzing the large $l$ behavior of the general term inside the summation in the corresponding contribution of ({\ref{TE}}). Here, with the propose to investigate a possible divergent contribution in the core-induced term, we shall adopt a similar expansion. Considering the solutions (\ref{RBP}), and after many intermediate steps, we found that for large value of $l$, the general term is proportional to $\frac{(a/r)^l}{l^2}$, whose summation provides a finite result for $r=a$. With this result we can conclude that there is no divergent contribution on the boundary for the self-energy in the ballpoint pen model.\footnote{This result is essentially different from the corresponding one obtained in \cite{Barbosa}, which presents a logarithmic divergent result near the boundary; however in the latter publication this singular result is not correct. An errata is being prepared by us which should be submitted to the respective journal.}

Let us now analyze the scalar self-energy for the region inside the monopole. For the ballpoint pen model we need to substitute the functions (\ref{RBP}) into the formulas (\ref{G0in}) and (\ref{Galfain}) to construct the corresponding Green function, and in (\ref{G0renin}) to obtain the renormalized self-energy, (\ref{ERin}). So we have:
\begin{equation}
G_{0}(\vec{r},\vec{r}')=\frac{1}{8\alpha\sqrt{rr'}}\sum_{l=0}^{\infty}(2l+1)(-1)^lP_\chi^{-l-1/2}(x_{r_<})P_\chi^{l+1/2}(x_{r_>})P_{l}(\cos \gamma ) \ ,  \label{G0inBP}
\end{equation}
where
\begin{eqnarray}
\label{X}
	x_{r_>}=\sqrt{1-\frac{r_{>}^{2}\sin^{2}\epsilon}{a^{2}}} \  \  {\rm and} \  \ x_{r_<}=\sqrt{1-\frac{r_{<}^{2}\sin^{2}\epsilon}{a^{2}}} \ .
\end{eqnarray}
As to the singular expansion of the Green function we use (\ref{Had3D}). 

Taking first the coincidence limit on the angular variables, and using the the coordinate system adopted in (\ref{gm2}) with (\ref{v}), the radial geodesic distance between to close points can be approximately given by,
\begin{eqnarray}
\sqrt{2\sigma}\approx\frac{\alpha|r'-r|}{\sqrt{1-(r/a)^{2}\sin^{2}\epsilon}} \ .
\end{eqnarray}
So we get
\begin{eqnarray}
\label{Gsing-BP}
G_{Sing}(r',r)=\frac{\sqrt{1-(r/a)^2\sin^{2}\epsilon}}{4\pi\alpha |r'-r|}-\frac{m}{4\pi} \ .
\end{eqnarray}

Developing some intermediate steps, explicitly presented in Appendix \ref{appB}, we obtain the following expression for $G_{0,Ren}(r',r)$:
\begin{eqnarray}
\label{Gr0}
G_{0,Ren}(r,r)=\frac1{4\pi\alpha r}S_\alpha(r/a)+\frac{m}{4\pi} \ ,
\end{eqnarray}
where
\begin{eqnarray}
\label{Gr11}
S_\alpha(r/a)=\sum_{l=0}^\infty\left[F\left( -\chi, \ \chi+1; \ 1/2-l; \ z_r/2\right)F\left( -\chi, \ \chi+1; \ 3/2+l; \ z_r/2\right)-1\right] \ ,
\end{eqnarray}
being
\begin{eqnarray}
\label{Gr2}
z_r=1-\sqrt{1-(r/a)^2\sin^2\epsilon} 
\end{eqnarray}
and $F(a,b;c;z)$ the hypergeometric function \cite{Grad}.

Consequently the renormalized self-energy becomes,
\begin{eqnarray}
\label{EintBP}
E_{Ren}=\frac{q^2}{8\pi\alpha r_p}S_\alpha(r_p/a)+\frac{q^2m}{8\pi}-\frac{q^2}{16\alpha r_p}\sum_{l=0}^\infty(2l+1)(-1)^lD_l^{(-)}(a) \left(P_\chi^{-l-1/2}(x_{r})\right)^2 \ .
\end{eqnarray}

For the special case where $m=0$ and $\xi=1/8$, we get $\chi=0$, consequently each hypergeometric function inside (\ref{Gr11}) becomes equal to unity providing a vanishing value to $S_\alpha(r/a)$. So all the contribution for the self-energy comes from the core-induced term. The  cancellation of the contribution due to the inner geometry can be directly verified by showing that the corresponding Green function, $G_{0}(\vec{r},\vec{r}')$, can be expressed in terms of the Green function in a flat three dimensional space. In order to prove that, we write the associated Legendre function in terms of hypergeometric one by \cite{Grad}:
\begin{eqnarray}
\label{Legend}
	P_\chi^\mu(x)=\frac1{\Gamma(1-\mu)}\left(\frac{1+x}{1-x}\right)^{\frac\mu2}F\left(-\chi, \ \chi+1; \ 1-\mu; \frac{1-x}2\right) \ .
\end{eqnarray} 
Taking $\chi=0$, the associated Legendre function above becomes an elementary function. Substituting this expression into (\ref{G0inBP}), taking $\gamma=0$ and using the relation for the Gamma functions, $\Gamma(3/2+l)\Gamma(1/2-l)=\frac{(2l+1)\pi(-1)^l}2$, we get,
\begin{eqnarray}
\label{GY}
G_0(r',r)=\frac{{\cal{Y}}^{1/4}}{4\pi\alpha\sqrt{r'r}}\sum_{l=0}^\infty{{\cal{Y}}}^{l/2}=\frac{{\cal{Y}}^{1/4}}{4\pi\alpha\sqrt{r'r}}\frac1{1-\sqrt{\cal{Y}}} \ ,
\end{eqnarray}
where
\begin{eqnarray}
\label{Y}
{\cal{Y}}=\frac{x_{r_<}-1}{x_{r<}+1}\frac{x_{r>}+1}{x_{r>}-1}	 \ .
\end{eqnarray}
Taking the limit $r'\to r$ we find,
\begin{equation}
G_0(r',r)\longrightarrow \frac{\sqrt{1-(r/a)^2\sin^{2}\epsilon}}{4\pi\alpha |r'-r|} \ .
\end{equation}

The space components of the metric tensor in the region inside the monopole is conformally related with the flat space by a conformal factor as shown below:
\begin{eqnarray}
d{\vec{l}}^2=v^2(r)dr^2+\alpha^2r^2d\Omega_{(2)}=f^2(\rho)(d\rho^2+\rho^2d\Omega_{(2)}) \ ,
\end{eqnarray}
where 
\begin{eqnarray}
f(\rho)=\frac{4\alpha}{1+4(\rho/a)^2\sin^2\epsilon} \ {\rm with} \ \rho=\frac{r}{2+2\sqrt{1-(r/a)^2\sin^2\epsilon}} \ .
\end{eqnarray}
Due to this conformal transformation, the Green function associated with massless field conformally coupled with the geometry, can be expressed in terms of the Green function in Minkowski space by $G_0(\vec{\rho},\vec{\rho}')=f^{-1/2}(\rho) G_M(\vec{\rho},\vec{\rho}')f^{-1/2}(\rho')$, as shown below:
\begin{equation}
\label{Grr}
G_0(\rho',\rho)=\frac1{16\alpha\pi}\frac1{|\rho'-\rho|}\sqrt{1+4(\rho/a)^2\sin^2\epsilon}\sqrt{1+4(\rho'/a)^2\sin^2\epsilon} \ .
\end{equation}

Expressing the variable $x_r=\sqrt{1-(r/a)^2\sin^2\epsilon}$ in terms of the new variable $\rho$, we get ${\cal{Y}}=\left(\rho/\rho'\right)^2$. Substituting this result into (\ref{GY}) and developing some intermediate steps we reobtain (\ref{Grr}).

Let us now to calculate the scalar self-energy near the core's center. Taking $r\to 0$, we may approximate $x_r\approx 1-(r^2/2a^2)\sin^2\epsilon$, and by (\ref{Legend}) the leading term in the associated Legendre function is,
\begin{eqnarray}
P_\chi^{-l-1/2}(x_r)\approx\frac1{\Gamma(l+3/2)}\left(\frac{r\sin\epsilon}{2a}\right)^{l+1/2} \ .
\end{eqnarray}
The dominant component of boundary induced term is for $l=0$. Considering also the contribution due to the mass term in (\ref{EintBP}) we have
\begin{equation}
E_{Ren}\approx-\frac{q^2D_0^{(-)}(a)\sqrt{1-\alpha^2}}{8\alpha\pi a}+\frac{q^2m}{8\pi} \ .
\end{equation}
Taking now the expression for $D_0^{(-)}(a)$ and considering $\xi=0$, we get:
\begin{eqnarray}
D_0^{(-)}(a)=\frac{P_\chi^{1/2}(\alpha)(1+ma)\alpha-P_{-\chi}^{1/2}(\alpha)}{P_\chi^{-1/2}(\alpha)(1+ma)\alpha} \ .
\end{eqnarray}
In the limit of vanishing mass \cite{Grad},
\begin{eqnarray}
D_0^{(-)}(a)=-\frac{\sqrt{1-\alpha^2}}\alpha \ ,
\end{eqnarray}
providing the self-energy below
\begin{eqnarray}
E_{Ren}\approx-\frac{q^2(\alpha^2-1)}{8\pi\alpha^2a} \ .
\end{eqnarray}

Before to finish this application we want to cal attention that, as the case of region outside the core, there is no divergent result given by core-induced part near the boundary.

In figure $4$ we exhibit the behavior of the renormalized self-energy for a charged test particle placed inside (left plot) and outside (right plot) the core as function of $mr$. In these analysis we consider minimal coupling ($\xi=0$), $\alpha=0.7$ and the product $ma=1$.\footnote{By our numerical analysis we observed that the value of the left and right plot at the point $mr=1$ are approximately $0.28$.} 
\begin{figure}[tbph]
\begin{center}
\label{fig4}
\begin{tabular}{cc}
\epsfig{figure=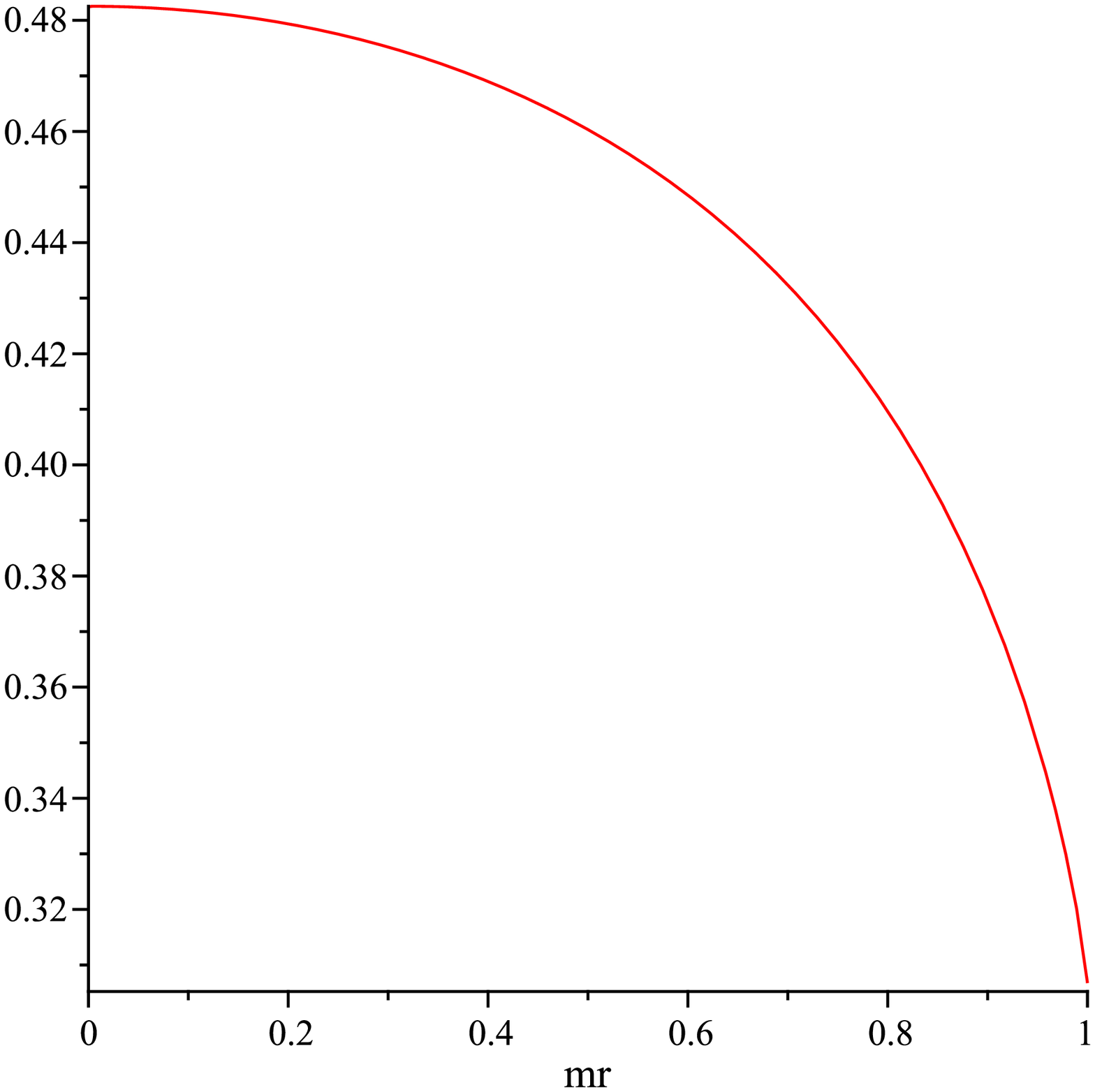, width=6.5cm, height=6.5cm,angle=0} & \quad 
\epsfig{figure=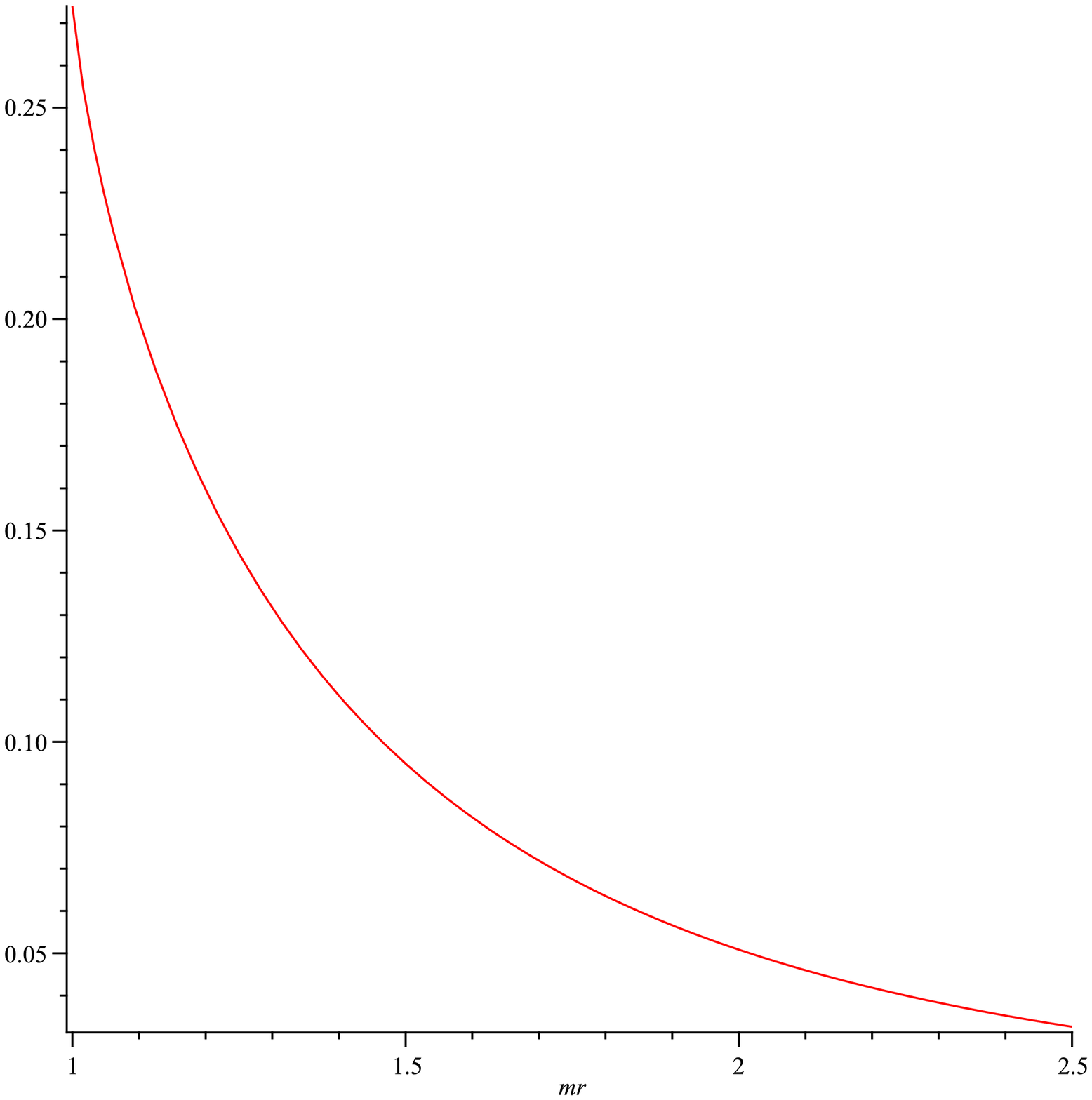, width=6.5cm, height=6.5cm,angle=0}
\end{tabular}
\end{center}
\caption{These graphs provide the behavior of $\frac{8\pi E_{Ren}}{q^2m}$ as function of $mr$ for the ballpoint pen model considering $\xi=0$ and $\alpha=0.7$. In the left plot we exhibit the behavior for the region inside the monopole, and in the right plot for the region outside the monopole.}
\label{fig4}
\end{figure}

\section{Concluding remarks}
\label{Conc}
One of the most interesting phenomenon associated with charged sources in the spacetimes which present non-trivial topology and/or curvature, is associated with the induced self-interactions on the sources themselves. In this sense the manifold modify the self-energy, when compared with the corresponding one in a Minkowski background. Although both self-energies are divergent, by analyzing the correction on this energy produced by the non-trivial topology and/or curvature of this spacetime, named renormalized self-energy, we observe that it is finite, exhibiting peculiar characteristic of the manifold under consideration.

In fact, the renormalized self-energies associated with static point-like charged test particles, have been investigated by many authors cited along of this paper for specific spacetimes which allow calculations of three-dimensional Green functions in closed forms. Here in this paper, we have added a new discussion about this subject, analyzing the self-energy associated with a point-like charged scalar test particle placed at rest in a global monopole spacetime considering a inner structure to it. In this way, the renormalized self-energy presents two contributions: one due to background geometry of spacetime where the particle is placed and the other induced by the boundary.

Because we have considered particle at rest, all equations found in this analysis became effectively defined in a three-dimensional space. Specifically the field equation (\ref{EM1}) is similar to the corresponding one obtained for an electric charged particle; the main differences reside in the presence of a mass term and a curvature coupling in that equation. Moreover, because the space section of the metric tensor in the regions inside and outside the monopole are conformally flat, the corresponding contributions to the scalar renormalized self-energies vanish for massless field conformally coupled with the geometry. For this case only the core-induced contribution remains.

As the main conclusions found in this work, three deserves to be mentioned. They are: $i)$ The renormalized self-energy depends strongly on the value adopted for the curvature coupling constant $\xi$. For specific values of this constant, the self-energy may provide repulsive, or attractive self-forces with respect to the boundary. $ii)$ The self-energy presents a finite value at the monopole's core center, and $iii)$ the behavior of the self-energy near the boundary depends also crucially on the junction condition obeyed by the metric tensor on it. By our results, for flower-pot model which presents Dirac-delta type contribution to the Ricci scalar, the core-induced part on the self-energy is logarithmically divergence near the boundary. As to the ballpoint-pen model, the self-energy is a continuous function of the radial coordinate $r$. Moreover, by numerical analysis not presented here, we can conclude that the induced self-force in this model is also a continuous function of $r$. The behavior of the self-energies in both models near the boundary, are exhibited numerically in figures $3$ and $4$.

\section*{Acknowledgment}
ERBM thanks CNPq. for partial financial support and FAPES-ES/CNPq. (PRONEX). 

\appendix
\section{The renormalization of the Green function $G_{gm}(\vec{r},\vec{r}')$}
\label{appA}
The expression for the Green function, $G_{gm}(\vec{r},\vec{r}')$, associated with scalar massive fields in a point-like global monopole spacetime is given in (\ref{G-gm}), which we reproduce below
\begin{eqnarray}
\label{A1}
G_{gm}(\vec{r},\vec{r}')=\frac1{4\pi\alpha^2{\sqrt{rr'}}}\sum_{l=0}^\infty (2l+1)I_{\nu_l}(mr_<)K_{\nu_l}(mr_>)P_l(\cos\gamma) \ ,
\end{eqnarray}
with
\begin{equation}
\nu_l=\frac1{2\alpha}\sqrt{(2l+1)^2+(1-\alpha^2)(8\xi-1)} \ .
\end{equation}
Unfortunately it is not possible to provide a closed expression for this Green function for arbitrary values of $\alpha$ and $\xi$. Moreover, because we are interested to analyze its behavior at coincidence limit, let us first take $\gamma=0$. In order to provide a well defined renormalized  expression for the above Green function at coincidence limit, let us divide the summation in (\ref{A1}) in two parts: $i)$ In the first part, $S_1$, we allow that the angular quantum number, $l$, goes from $0$ to $\bar{l}-1$, and $ii)$ in the second part, $S_2$, we allow that $l$ goes from $\bar{l}$ until infinity:
\begin{eqnarray}
\label{Ss}
S=\sum_{l=0}^\infty (2l+1)I_{\nu_l}(\rho_<)K_{\nu_l}(\rho_>)=S_1+S_2 \ ,
\end{eqnarray}
where we have defined the dimensionless variable $\rho=mr$.

In the following analysis we shall consider $\bar{l}$ large enough which allow us to use in $S_2$ the asymptotic expansion for the modified Bessel functions \cite{Abra}:
\begin{eqnarray}
K_\nu(\nu z)&\approx&\sqrt{\frac{\pi}{2\nu}}\frac{e^{-\nu\eta(z)}}{(1+z^2)^{1/4}}\sum_{k=0}^\infty(-1)^k\frac{u_k(w(z))}{\nu^k} \ , \nonumber\\
\ I_\nu(\nu z)&\approx&\frac1{\sqrt{2\pi\nu}}\frac{e^{\nu\eta(z)}}{(1+z^2)^{1/4}}\sum_{k=0}^\infty\frac{u_k(w(z))}{\nu^k} \ ,
\end{eqnarray}
with $w(z)=\frac1{\sqrt{1+z^2}}$ and $\eta(z)=\sqrt{1+z^2}+\ln\left(\frac{z}{1+{\sqrt{1+z^2}}}\right)$. The relevant expressions for the functions $u_k(z)$ are: $u_0(w)=1$ and $u_1(w)=\frac{3w-5w^3}{24}$. In our analysis $z=\frac{\rho}{\nu}$. As to the order of Bessel function, $\nu_l$, we may expand it in powers of $\frac1{2l+1}$, and keep only the two firsts terms:
\begin{equation}
\label{nu-aprox}
\nu_l\approx\frac{(2l+1)}{2\alpha}+\frac{(1-\alpha^2)(8\xi-1)}{4\alpha(2l+1)}+O\left(\frac1{(2l+1)^2}\right) \ .
\end{equation}
After some intermediate steps we get:
\begin{eqnarray}
\label{IK-aprox}
I_{\nu_l}(\rho_<)K_{\nu_l}(\rho_>)\approx\frac{t^{\nu_l}}{2\nu_l}+\frac{t^{\nu_l}}{2\nu_l^2}\left(\frac{\rho_<^2-\rho_>^2}{4}+u_1(w_<-u_1(w_>)\right)+O\left(\frac1{\nu_l^3}\right)  \ ,
\end{eqnarray}
where $t=\frac{r_<}{r_>}$. Taking the approximated expression (\ref{nu-aprox}), we have
\begin{eqnarray}
\label{t-aprox}
t^{\nu_l}\approx t^{(2l+1)/(2\alpha)}+\ln(t)\frac{(1-\alpha^2)(8\xi-1)}{4\alpha}\frac{t^{(2l+1)/(2\alpha)}}{(2l+1)}+ \ ...
\end{eqnarray}

Now substituting (\ref{IK-aprox}) and (\ref{t-aprox}) into the second summation of (\ref{Ss}), we obtain:
\begin{eqnarray}
S_2&\approx& \alpha t^{1/(2\alpha)}\frac{t^{{\bar{l}}/\alpha}}{1-t^{1/\alpha}}+\ln(t)t^{1/(2\alpha)}\frac{(1-\alpha^2)(8\xi-1)}{4} \frac{t^{{\bar{l}}/\alpha}}{2} \Phi(t^{1/\alpha},1;\bar{l}+1/2)\nonumber\\
&-&\alpha^2m^2r_>^2t^{1/(2\alpha)}(1-t^2)\frac{t^{{\bar{l}}/\alpha}}{4}\Phi(t^{1/\alpha},1;\bar{l}+1/2)+... \ ,
\end{eqnarray} 
where $\Phi(z,s,v)$ is the Lerch Phi function \cite{Pru}.\footnote{The Lerch Phi function is defined as $\Phi(z,s,v)= \sum_{n=0}^\infty\frac{z^n}{(v+n)^s}$.} 

We can see that only the first term in the above expression diverges in the limit, $t\to 1$. In fact all the other terms vanish at this limit, and the omitted ones provide finite results.

Now going back to the summation $S$, we may write:
\begin{eqnarray}
S&\approx&\sum_{l=0}^{\bar{l}-1} (2l+1)I_{\nu_l}(\rho_<)K_{\nu_l}(\rho_>)+\alpha t^{1/(2\alpha)}\frac{t^{{\bar{l}}/\alpha}}{1-t^{1/\alpha}}+ \ln(t)t^{1/(2\alpha)}\frac{(1-\alpha^2)(8\xi-1)}{4} \frac{t^{{\bar{l}}/\alpha}}{2}\nonumber\\
&&\times \Phi(t^{1/\alpha},1;\bar{l}+1/2)-\alpha^2m^2r_>^2t^{1/(2\alpha)}(1-t^2)\frac{t^{{\bar{l}}/\alpha}}{4}\Phi(t^{1/\alpha},1;\bar{l}+1/2)+... \ .
\end{eqnarray}

The renormalized Green function is written in terms of $S$ by,
\begin{eqnarray}
G_{gm,ren}=\frac1{4\pi\alpha r_>}\left\{t^{-1/2}\frac{S}\alpha-\frac{\alpha}{1-t}\right\}+\frac{m}{4\pi} \ .
\end{eqnarray}
Substituting $S$ into the above function and taking the coincidence limit, we obtain\footnote{In all the calculations we have discarded th summation over the terms of order equal or bigger than $\frac1{\nu_l^2}\approx\frac1{(2l+1)^2}$. The corresponding summations in the limit $t=1$, provide $\Psi(n,{\bar{l}})$ which is the nth polygamma function. These functions go to zero for ${\bar{l}}\to \infty$.} 
\begin{eqnarray}
G_{gm,ren}\approx\frac1{4\pi\alpha r}\sum_{l=0}^{\bar{l}-1}\left[\frac{(2l+1)}\alpha I_{\nu_l}(mr)K_{\nu_l}(mr)-1\right]+\frac{m}{4\pi}+... \ .
\end{eqnarray}

An exact result is obtained by taking ${\bar{l}}\to\infty$, which gives the result (\ref{Gmren2}).

\section{The renormalization of the Green function $G_0(r',r)$ in ballpoint pen model}
\label{appB}
The expression for the Green function, $G_0(r',r)$, in the region inside the monopole, considering the ballpoint pen model, is  
\begin{equation}
\label{g00}
G_{0}(\vec{r},\vec{r}')=\frac{1}{8\alpha\sqrt{rr'}}\sum_{l=0}^{\infty}(2l+1)(-1)^lP_\chi^{-l-1/2}(x_{r_<})P_\chi^{l+1/2}(x_{r_>})P_{l}(\cos \gamma ) \ .
\end{equation}
Taking $\gamma=0$ and using (\ref{Legend}), we can express the term inside the summation above by
\begin{eqnarray}
\label{Y}
{\cal{Y}}^{\frac l2+\frac14}F\left( -\chi, \ \chi+1; \ 1/2-l; \ (1-x_{r_>})/2\right)F\left( -\chi, \ \chi+1/2; \ 3/2+l; \ (1-x_{r_<})/2\right) \ ,
\end{eqnarray}
with
\begin{eqnarray}
{\cal{Y}}=\frac{x_{r_<}-1}{x_{r<}+1}\frac{x_{r>}+1}{x_{r>}-1}	 \ ,
\end{eqnarray}
being $x_{r_>}$ and $x_{r_<}$ given in (\ref{X}). In order to observe the divergent contribution of $G_0$, let us analyze (\ref{Y}) in the limit of large value for $l$. Using the series expansion for the hypergeometric function, the corresponding behavior is
\begin{eqnarray}
	\left[{\cal{Y}}^{\frac l2}-\frac{\chi(\chi+1)}2(x_{r_>}-x_{r_<})\frac{{\cal{Y}}^{\frac l2}}l+O\left(\frac1{l^2}\right)\right]{\cal{Y}}^{1/4} \ .
\end{eqnarray}
The contributions of the first and second terms above in the summation in (\ref{g00}) are, respectively, $\frac1{1-\sqrt{\cal{Y}}}$ and $-\frac{\chi(\chi+1)}2(x_{r_>}-x_{r_<})\ln(1-\sqrt{\cal{Y}})$. However in the coincidence limit the second contribution provide a vanishing result. Considering  the limit $x_{r_<}\to x_{r_>}$, we have
\begin{eqnarray}
	{\cal{Y}}\approx 1-2\frac{x_{r_>}-x_{r_<}}{x_{r_<}^2-1}+O(x_{r_>}-x_{r_<})^2 
\end{eqnarray}
and consequently 
\begin{eqnarray}
	1-\sqrt{\cal{Y}}\approx\frac{r_>-r_<}{r_>\sqrt{1-\left(\frac{r_<}{a}\right)^2\sin^2\epsilon}} \ .
\end{eqnarray}
Finally the contribution due to this term in (\ref{g00}) is
\begin{eqnarray}
	G_0(r',r)=\frac1{4\pi\alpha r}\frac{\sqrt{1-(r_</a)^2\sin^2\epsilon}}{1-r_</r_>}+ \  ....
\end{eqnarray}
which coincides with the first term in (\ref{Gsing-BP}). Finally the renormalized Green function is given by 
\begin{eqnarray}
	G_{0,ren}(r,r)&=&\frac1{4\pi\alpha r}\lim_{r'\to r}\sum_{l=0}^\infty\left[{\cal{Y}}^{\frac l2+\frac14}F\left( -\chi, \ \chi+1; \ \ 1/2-l; \ (1-x_{r_>})/2\right)\times\right.\nonumber\\
&&\left.F\left( -\chi, \ \chi+1; \ 3/2+l; \ (1-x_{r_<})/2\right)-{\cal{Y}}^{\frac l2}\right]+\frac{m}{4\pi}\nonumber\\
&=&\frac1{4\pi\alpha r}S_\alpha(r/a)+\frac{m}{4\pi} \ .
\end{eqnarray}

\end{document}